\documentclass{aa}
\usepackage[varg]{txfonts}
\usepackage{natbib}

\newcommand{\vect}[1]{\boldsymbol{#1}}
\renewcommand{\thesubsection}{\arabic{subsection}}
\usepackage{makecell}
\usepackage{hyperref}
\usepackage{verbatim}
\usepackage{pbox}
\usepackage{physics}

\newcommand{\mpe}{Max Planck Institute for extraterrestrial Physics, Giessenbachstr., 85748 Garching, Germany, \email{idelw@mpe.mpg.de}}
\newcommand{\ipag}{Univ. Grenoble Alpes, CNRS, IPAG, F-38000 Grenoble, France}

\begin{document}

\title{The Collimated Radiation in SS 433 \thanks{Based on observations collected at the European Southern Observatory, Chile, Program ID 099.D-0666(A,B)}}
\subtitle{Constraints from Spatially Resolved Optical Jets \\ and \texttt{Cloudy} Modeling of the Optical Bullets} 

\author{Idel Waisberg\inst{1} \and Jason Dexter\inst{1} \and Pierre Olivier-Petrucci\inst{2} \and Guillaume Dubus\inst{2} \and Karine Perraut\inst{2}} 
\institute{\mpe \and \ipag}

\abstract {The microquasar SS 433 is well-known for its precessing, relativistic and highly collimated baryonic jets, which manifest in its optical spectrum as pairs of hydrogen and helium emission lines moving across with large Doppler shifts. Depending on their heating mechanism, the optical jet bullets may serve as a probe of the collimated radiation coming from the inner region close to the compact object, and which is not directly visible to observers on Earth.}
{We aim to better understand the baryonic jet phenomenon in SS 433, in particular the properties of the optical bullets and their interaction with the ionizing collimated radiation.} 
{The optical interferometer VLTI/GRAVITY has allowed to spatially resolved the optical jets in SS 433 for the first time. We present here the second such observation taken over three nights in July 2017. In addition, we use the multi-wavelength XSHOOTER spectrograph at VLT to study the optical bullets in SS 433 in detail. Over the full wavelength range $0.3-2.5 \mu$m, we identify up to twenty pairs of jet lines observed simultaneously, which we model with the spectral synthesis code \texttt{Cloudy}.}
{GRAVITY reveals elongated exponential-like spatial profiles for the optical jets, suggestive of a heating mechanism acting throughout a long portion of the jet and naturally explained by photoionization by the collimated radiation. We also spatially resolve the movement of the optical bullets for the first time, detecting extended jet components corresponding to previous ejections. \texttt{Cloudy} photoionization models explain both the spatial intensity profiles measured with GRAVITY and the line ratios from XSHOOTER, and constrain the properties of the optical bullets and the ionizing radiation. We find that the latter must peak in the UV with an isotropic luminosity (as inferred by a face on observer) $\approx 10^{41}$ erg/s. Provided that the X-ray SED is sufficiently hard, the collimated X-ray luminosity could still be high enough so that the face on observer would see SS 433 as an ULX ($L_X \lesssim 10^{40}$ erg/s) and it would still be compatible with the H/He/He+ ionization balance of the optical bullets. The kinetic power in the optical jets is constrained to $3-20 \times 10^{38}$ erg/s, and the extinction in the optical jets to $A_V = 6.7 \pm 0.1$. We suggest there may be substantial $A_V \gtrsim 1$ and structured circumstellar extinction in SS 433, likely arising from dust formed in equatorial outflows.} 
{}


\maketitle

\section{INTRODUCTION}
\label{introduction}

After more than forty years since its discovery, SS 433 \citep{Stephenson77,Clark78} remains a unique object primarily due to its 
relativistic, precessing baryonic jets. They were first discovered through broad emission lines of hydrogen and helium 
moving across its optical spectrum with extreme Doppler shifts \citep{Margon79}. Their blue/redshifts follow a kinematic precession model 
\citep{Fabian79,Margon84}, according to which the jets precess over a period $\approx 163$ days, following a cone of half opening angle $\approx 21^\circ$ and 
a precessional axis that is inclined $\approx 78^\circ$ relative to the line of sight, with jet material moving at a very stable speed $\approx 0.26c$ along radial ballistic trajectories \citep{Eikenberry01}. 
From the width of the emission lines and assuming a conical geometry, the optical jets are very collimated, with half opening angle $\theta \lesssim 1^\circ$ \citep{Borisov87}. The jet precession is thought to be 
driven by gravitational torques by the compact object on the donor star \citep[slaved disk model; ][]{Roberts74,vandenHeuvel80,Whitmire80}, whose spin axis is 
misaligned with the orbital plane presumably due to the supernova explosion that originated the compact object $\lesssim 10^5$ years ago \citep{Zealy80,Lockman07}. SS 433 is also an eclipsing binary, 
and the behavior of the X-ray and optical eclipses together with radial velocities of the emission lines reveal that the supercritical disk and its outflows dominate the continuum and line radiation at all wavelengths 
\citep[for a review of the fantastic properties of SS 433, see][]{Fabrika04}. 

The jets in SS 433 were soon also detected in radio \citep{Hjellming81}, confirming the kinematic precession model derived from the optical lines and 
establishing the position angle of the precessional axis on sky $98.2^\circ$ \citep{Stirling02}. VLBI observations have regularly imaged the movement 
of individual jet knots \citep[e.g.][]{Vermeulen93,Paragi99}, and larger scale images of the corkscrew structure created by the precessing jets have established a precise distance 
$d = 5.5 \pm 0.2$ kpc from the aberration induced by the light travel time effect between the two jets \citep{Blundell04}. Later the jets were detected in X-rays \citep{Watson86}, where they are seen as emission lines from highly ionized metals (e.g. Fe, Ni, S, Si) which follow the same kinematic precession model as the optical jets and have similarly small opening angle \citep[e.g.][]{Kotani96,Marshall02}. They have been modeled with multi-temperature, optically thin collisional ionization models to estimate properties such as temperature, density and kinetic power. Because the X-ray jets are continuous, emission-line diagnostics depend on the assumptions on the geometry of the outflow, usually taken to be a radially outflowing cone, and may also be affected by photoionization from putative collimated radiation \citep{Brinkmann00}. The behavior of the X-ray jets during eclipse constrain their length $\gtrsim 10^{12} \text{ cm}$ \citep{Marshall13}.

Because of its edge-on orientation, little is known about the radiation in the beam containing the optical jets, nor about the radiation from the inner parts of the accretion flow in general. The supercritical, geometrically and optically thick disk is thought to reprocess the latter to a large radius \citep[$\sim R_{sp}$, the spherization radius within which radiation pressure leads to a thick disk,][]{Shakura73}, thermally downgrading it to the observed blackbody temperature $T \sim 30000$ K \citep{Fabrika04}. As a result, SS 433 is a relatively faint X-ray source $L_X \sim 10^{35} - 10^{36}$ erg/s to observers on Earth, with most of the received X-ray flux originating from isotropic thermal Bremststrahlung from the X-ray jets, without any apparent X-ray accretion disk \citep{Watson86}. It has been proposed that, if viewed face on, so that the inner portions of the accretion disk/jet funnel were directly visible, SS 433 would appear as an extremely bright X-ray source such as an ULX \citep{Fabrika04,Begelman06}. Recent optical spectroscopy of ULX counterparts have shown strong emission lines akin to those seen in SS 433 and likely associated with supercritical disks \citep{Fabrika15}, as well as the discovery of an Ultraluminous Supersoft Source (ULS) containing a baryonic relativistic jet seen in a moving H$\alpha$ line \citep{Liu15}, so far the only other known object to show such a feature. It remains an open question whether a subpopulation of ULXs could be SS433-like objects, or whether SS 433 is an even rarer phenomenon.  

The optical bullets that make up the optical jets are thought to form from the collapse of gas from the continuous X-ray jet through a thermal instability as the jet expands and cools \citep{Davidson80,Brinkmann88}. If the optical bullets are heated mainly by photoionization \citep{Bodo85,Fabrika87,Panferov93}, they can serve as a probe of the collimated radiation. On the other hand, their heating has generally been ascribed to external processes related to interaction with the ambient gas, either by direct collisions \citep{Davidson80,Brown91} or through shocks with subsequent heating and ionization \citep{Begelman80}. Spatially resolving the optical jets could reveal the dominant heating process; just the same, the optical jets have been associated with scales $\sim 10^{14} - 10^{15} \text{ cm} \leftrightarrow 1-10 \text{ mas}$ \citep[e.g.][]{Borisov87,Marshall13}, beyond the reach of current diffraction limited single telescopes. 

The only way to spatially resolve the optical jets of SS 433 is through interferometry. In \cite{GRAVITYSS43317} (Paper I) we presented the first such observations taken during commissioning of the GRAVITY instrument \citep{GRAVITY17} in July 2016 at the Very Large Telescope Interferometer (VLTI), which works in the near-infrared K band. These observations revealed that the optical jets peak very close to the binary and follow an extended exponential radial emission profile with decay constant $\approx 2$ mas, suggestive of a continuous heating process throughout the jet. Here, we present a second set of three GRAVITY observations of SS 433 taken over four nights in July 2017 in which we could observe the change in spatial emission profiles of the jets as the emission lines bright and fade. We also present the first XSHOOTER observations of SS 433, where we use up to twenty pairs of jet lines to constrain the properties of the bullets and the ionizing collimated radiation under the assumption of heating by photoionization suggested by the GRAVITY data. 

This paper is organized as follows. In Section 2, we summarize the observations and data reduction. The GRAVITY and XSHOOTER data analysis are presented in Sections 3 and 4, respectively. Finally, Section 5 contains the conclusions. 

We often quote the results in mas since that is the actual measured unit in interferometry. For convenience, we quote $1 \text{ mas} \leftrightarrow 8.2 \times 10^{13} \text{ cm} = 1180 R_{\odot} = 5.5 \text{ AU}$, assuming a distance $d = 5.5 (\pm 0.2)$ kpc \citep{Blundell04}. The GAIA DR2 distance $4.6 \pm 1.3 \text{ kpc}$ \citep{Luri18} is consistent with this value.

\section{OBSERVATIONS AND DATA REDUCTION} 
\label{Observations and Data Reduction}

\subsection*{GRAVITY} 

SS 433 ($K\approx8$) was observed with GRAVITY \citep{GRAVITY17} with the Unit Telescopes (UT) on VLTI on three nights over a period of four days in July 2017. Half of the K band light of SS 433 itself was directed to the fringe tracker (FT), which operates at $> 1000$ Hz to stabilize the fringes in the science channel (SC), allowing coherent integration over detector integration times of 10s in high spectral resolution ($R \approx 4000$). The FT operates in low resolution ($R\approx 20$) with five channels over the K band. The data were obtained in split polarization mode. The adaptive optics (AO) was performed at visual wavelength using SS 433 itself as the AO guide star ($V \approx 14$). 

Table \ref{table:phases} shows the precessional $\phi_{prec}$ and orbital $\phi_{orb}$ phases of each observation based on the ephemerides in \cite{Eikenberry01} and \cite{Goranskii98}, respectively. For more details on the observations and data reduction, including the uv coverage of the observations, we refer to the companion paper on the equatorial outflows (Waisberg et al., sub.). In light of a slightly improved jet model, we also reanalyze the 2016 observation (Paper I), which is also shown in Table \ref{table:phases}.  

\begin{table}[t]
\centering
\caption{Summary of Observations}
\label{table:phases} 
\begin{tabular}{cccc}
\hline \hline 
Date & Instrument & $\phi_{prec}$\tablefootmark{a} & $\phi_{orb}$\tablefootmark{b} \\[0.3cm]
\shortstack{2016-07-17\\(Paper I)} & GRAVITY & 0.71 & 0.11 \\[0.3cm]
\shortstack{2017-05-21\\Epoch X1} & XSHOOTER & 0.61 & 0.67 \\[0.3cm]
\shortstack{2017-05-28\\Epoch X2} & XSHOOTER & 0.65 & 0.21 \\[0.3cm] 
\shortstack{2017-06-20\\Epoch X3} & XSHOOTER & 0.79 & 0.97 \\[0.3cm] 
\shortstack{2017-06-30\\Epoch X4} & XSHOOTER & 0.85 & 0.72 \\[0.3cm]
\shortstack{2017-07-07\\Epoch 1} & GRAVITY & 0.895 & 0.25 \\[0.3cm]
\shortstack{2017-07-09\\Epoch 2} & GRAVITY & 0.907 & 0.40 \\[0.3cm]
\shortstack{2017-07-10\\Epoch 3} & GRAVITY & 0.913 & 0.48 \\[0.3cm]
\shortstack{2017-07-15\\Epoch X5} & XSHOOTER & 0.94 & 0.87 \\[0.3cm] 
\hline
\end{tabular}
\tablefoot{
\newline
\tablefoottext{a}{Based on the kinematic parameters in \cite{Eikenberry01}. Phase zero is when the eastern/western jet is maximally blueshifted/redshifted.} \newline
\tablefoottext{b}{Based on the orbital parameters in \cite{Goranskii98}. Phase zero corresponds to the eclipse center of the accretion disk.} 
}
\end{table}

\subsection*{XSHOOTER} 

SS 433 was observed five times between May and July of 2017 with the XSHOOTER \'echelle spectrograph \citep{Vernet11} mounted 
on the Very Large Telescope (VLT) at ESO in Paranal, Chile. The exposure times per epoch were 1240s, 1260s and 1248s 
for the UVB ($0.3 - 0.55 \mu$m), VIS ($0.55 - 1.0 \mu$m) and NIR ($1.0-2.5 \mu$m) arms. The slit dimensions
were 1.0"$\times$11", 0.9"$\times$11" and 0.4"$\times$11" for each arm, corresponding to spectral resolutions 
$R = \frac{\lambda}{\Delta \lambda}$ of 5400, 8900 and 11600. The observations were made in nodding pattern 
for sky subtraction. Table \ref{table:phases} shows the precessional $\phi_{prec}$ and orbital $\phi_{orb}$ phases of each observation. 

The data were reduced with the standard ESO X-Shooter pipeline (version 2.9.0), which includes de-biasing, flat-fielding, wavelength 
calibration, sky subtraction, order merging and flux calibration, the latter based on nightly response curves from flux standard stars.  
The line-modeling software Molecfit \citep{Smette15, Kausch15} was 
used to correct for telluric absorption in the VIS and NIR arms. The performance of Molecfit was found to be at least as good as 
using a telluric calibrator star, with the additional benefit that manual removal of the many H I, He I and additional lines of a telluric calibrator is not 
needed.

Although our observations were not designed for precise flux calibration, the latter has to be taken into account when comparing emission line strengths across 
a large wavelength range since SS 433 has a complex and variable continuum. Because the slit widths used are smaller than the seeing, it is necessary to 
correct for the wavelength-dependent slit losses. We have done this by assuming the typical wavelength dependence for seeing $s \propto \lambda^{-0.2}$, 
and using the overlapping regions between the spectral arms to fit for the average seeing, which is in agreement with the estimated value from the acquisition 
image at the start of each observation. 

\section{GRAVITY DATA ANALYSIS}
\label{GRAVITY} 

\subsection*{The K band Spectrum} 

\begin{figure*}[tb]
\centering
\includegraphics[width=2.0\columnwidth]{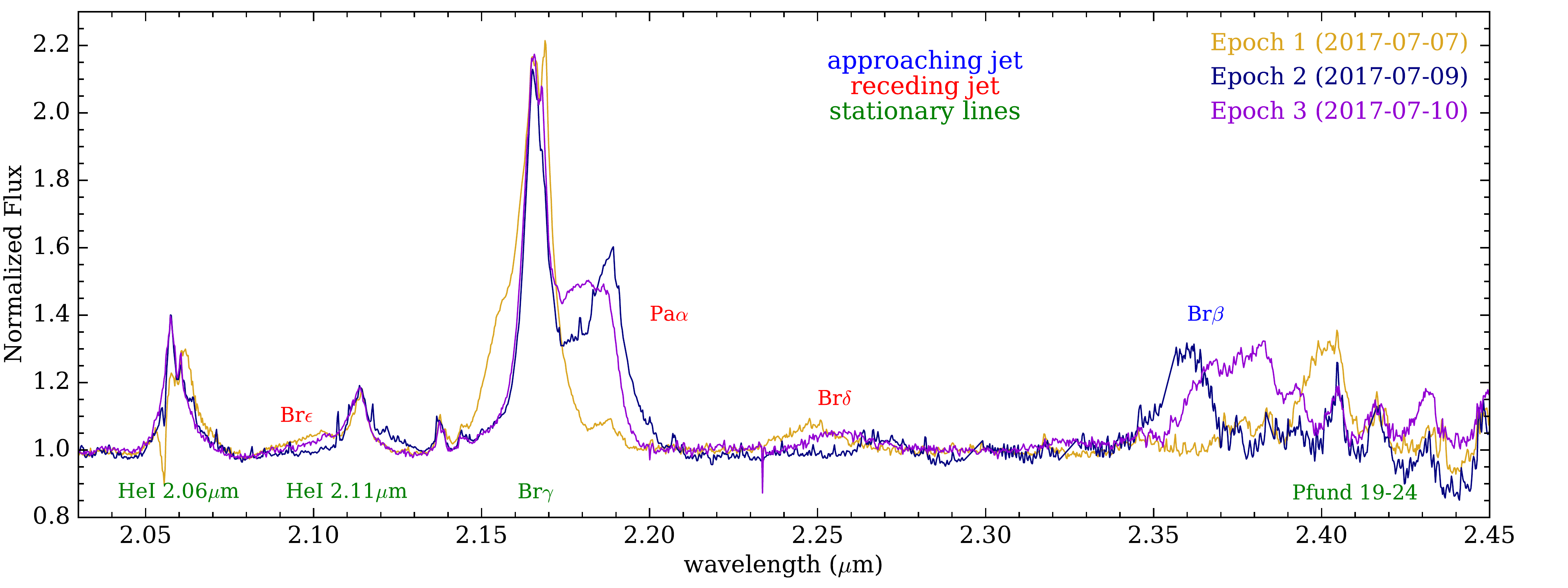}
\caption{K band spectra for the three GRAVITY observations in 2017. The strongest jet emission lines are Pa$\alpha$ and Br$\beta$ for the receding and approaching jets, respectively (the former is blended with the Br$\gamma$ stationary line, the latter with stationary Pfund lines). The brightening and fading of different jet knots are visible over the three observations. Stationary emission lines are labeled in green, but are not the subject of this paper.}
\label{fig:spectrum} 
\end{figure*}

Figure \ref{fig:spectrum} shows the GRAVITY spectra of SS 433 for the 2017 observations. There are stationary emission lines (Br$\gamma$, He I 2.06 $\mu$m, He I 2.11 $\mu$m and high order (upper levels 19-24) Pfund lines) as well as emission lines from the baryonic jets. For the analysis of the stationary Br$\gamma$ line, we refer to the companion paper on the equatorial outflows (Waisberg et al., sub.). In this paper, we focus on the jets. In the 2016 observation (Paper I), the precessional phase was such ($\phi_{prec} \approx 0.7$) that jet lines from Br$\gamma$, Br$\delta$ and He I 2.06 $\mu$m fell in the K band spectrum (see Figure A.1). In the 2017 observations presented here, the precessional phase was significantly different ($\phi_{prec} \approx 0.9$) so that other jet lines were observed: Br$\beta$ from the approaching jet and Pa$\alpha$ from the receding jet (as well as very weak Br$\delta$ and Br$\epsilon$ lines from the receding jet). Another difference in the latter observations is that there are often two components (knots) to the jet lines. The set of three observations over four nights allows to follow the spatial evolution of the jets; while the knots in Epoch 1 had faded by Epoch 2, the knots in Epochs 2 and 3 partially overlap. The jet redshifts were such that the Pa$\alpha$ jets are partially blended with the Br$\gamma$ stationary line and the Br$\beta$ jets with stationary Pfund lines. 
 
\begin{figure}[tb]
\centering
\includegraphics[width=\columnwidth]{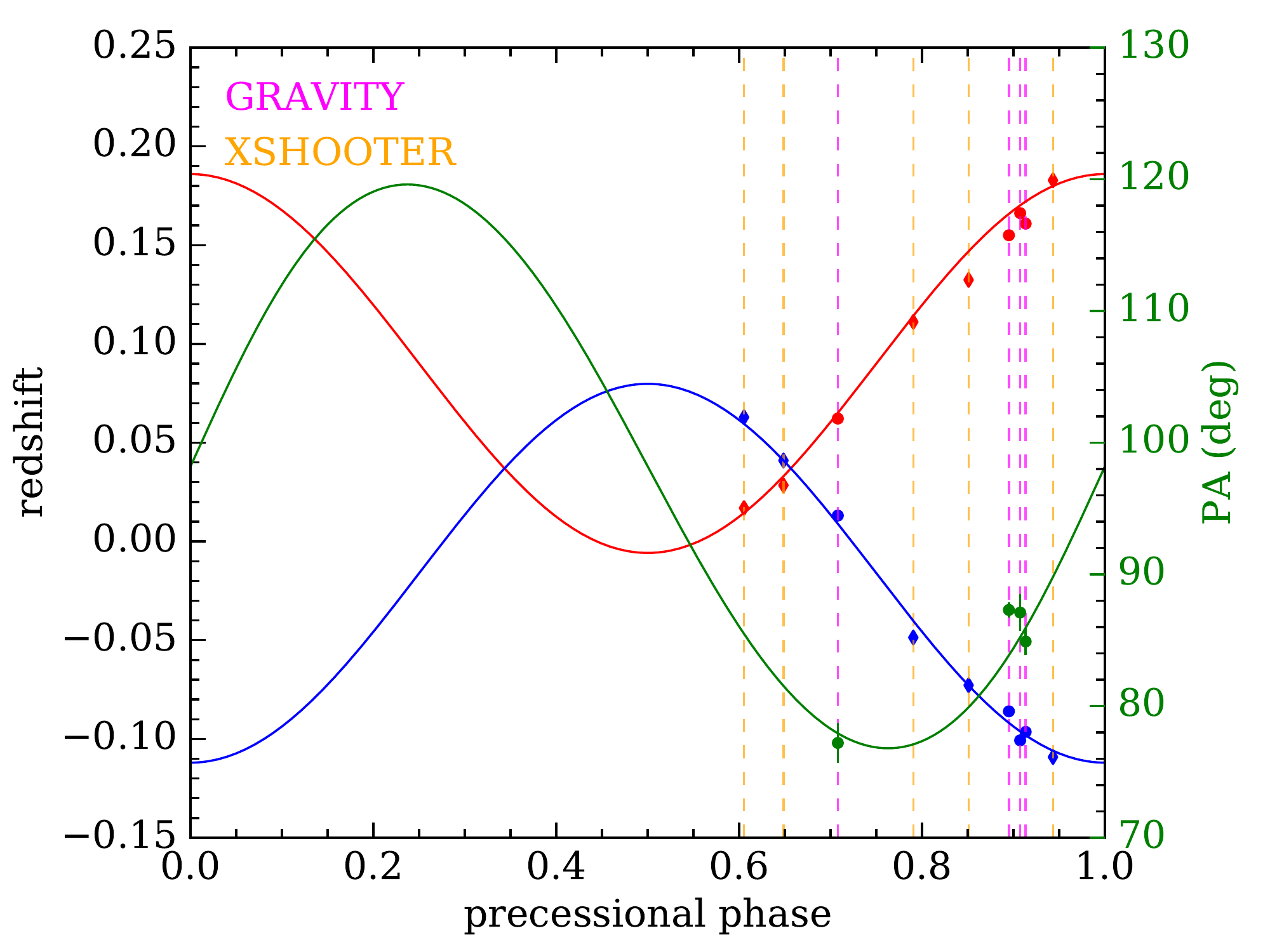} \\ 
\caption{Jet redshifts (mostly approaching (eastern) jet in blue; mostly receding (western) jet in red) and position angle on sky (green) of the optical jets for each observation as a function of precessional phase. The curves were made using the kinematic model parameters in \cite{Eikenberry01} and the precessional axis position angle in \cite{Stirling02}. The PA of the optical jets as derived from the GRAVITY data agrees well with the prediction from the radio jets. We plot both the 2016 ($\phi_{prec} \approx 0.7$) as well as the 2017 ($\phi_{prec} \approx 0.9$) GRAVITY observations.}
\label{fig:redshift}
\end{figure}

Figure \ref{fig:redshift} shows the measured jet redshifts along with the kinematic precession model using the parameters from \cite{Eikenberry01}. They agree within the perturbations caused by nutational motion and random jitter \citep{Fabrika04}. 

\subsection*{Optical Jet Interferometric Model} 

\begin{figure*}[tb]
\label{fig:model_plot_P99_3} 
\centering
\includegraphics[width=2.0\columnwidth]{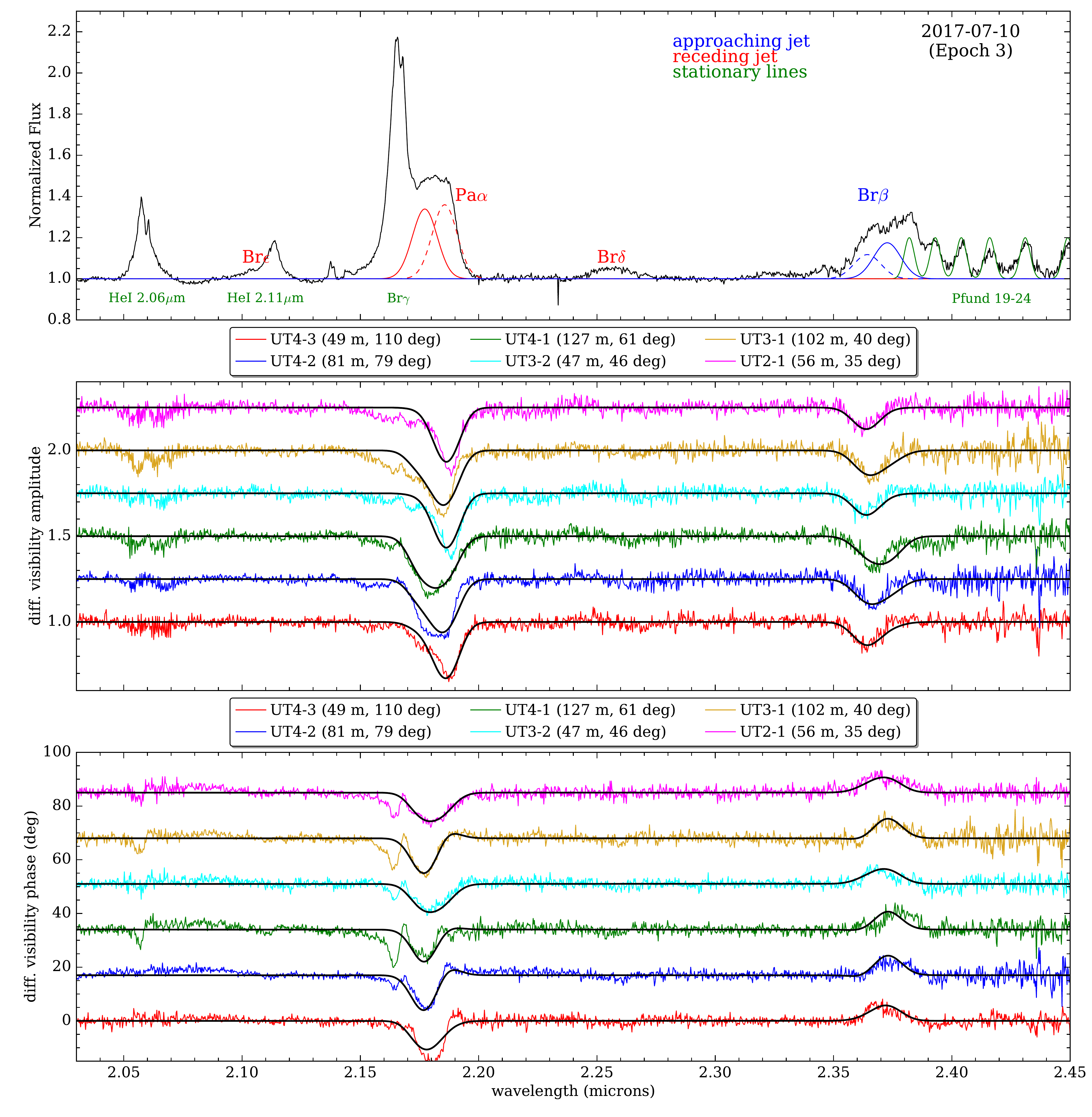}
\caption{Data and best fit jet model for the 2017 Epoch 3 GRAVITY observation. Compact (recent) and extended (older) jet knots are shown in full and dashed lines, respectively. Stationary lines are labeled in green. There is substantial blending of the Pa$\alpha$ lines from the receding jet with the stationary Br$\gamma$ line, which also has differential visibility signatures across it. The fits are done to all lines simultaneously, but we plot only the model for jet lines for clarity. The baselines projected lengths and position angles on sky are indicated in the labels.}
\end{figure*}

Figure \ref{fig:model_plot_P99_3} shows the differential visibility amplitudes and phases for Epoch 3 of the 2017 observations. The only strong jet lines present in the GRAVITY spectrum are Pa$\alpha$ for the receding jet and Br$\beta$ for the approaching jet. As in Paper I, the jets line emitting regions are more extended than the region emitting near-infrared continuum as shown by the decrease of the visibility amplitude across the lines, and the receding and approaching jets have opposite differential visibility phases (suggesting that they are on opposite sides) relative to the continuum. 

We fit the jet lines and differential visibilities simultaneously. For the spectrum, we fit the jet knots as Gaussians. The model parameters are:

\begin{enumerate}

\item The jet inclination $i$, which determines the redshift via 

\begin{align}
z = \gamma (1 \pm \beta \cos(i) )
\end{align}

\noindent where $\beta = 0.26$ is the jet speed in units of $c$ and $\gamma = \dfrac{1}{\sqrt{1-\beta^2}}$. Because we can measure the redshift very precisely for each jet line, we allow for different inclinations of the receding and approaching jets, as well as the different components. We assume a constant velocity since it is very stable \citep{Eikenberry01}, and the possible small variations in velocity are absorbed in the inclination. 

\item The FWHM in km/s, which is assumed to be the same for all the jet lines for all components; 

\item The strength of each jet line relative to the normalized continuum.

\end{enumerate}

To fit the visibilities, we model the jets as a 1D structure with a radial emission profile, since they are very collimated 
(opening angle $\lesssim 1^{\circ}$) from the width of the optical emission lines \citep{Borisov87}. The model parameters are:

\begin{enumerate}

\item The position angle (PA) of jet in the sky plane. It is assumed to be the same for all the jet components in a given epoch because it cannot be measured nearly as precisely as the 
inclination (the typical error is a few degrees); 

\item The radial emission profile is controlled by three parameters $\theta$, $\alpha$ and $r_0$: 

\begin{equation}
I (r) = \begin{cases} 
	r^{\alpha-1} e^{-r/\theta} & r \geq r_0 \\
	0 & r < r_0 
	\end{cases} 
\end{equation}

\noindent where $r$ is the distance from the center. This model is similar to the one used in Paper I, except for the 
additional parameter $\alpha$ which allows for more general shapes besides an exponential ($\alpha=1$), from Gaussian-like 
($\alpha > 1$) to steeper profiles ($\alpha < 1$). The parameters $\theta$ (which together with $\alpha$ controls the shape of the profile) 
and $r_0$ (the inner edge where the emission starts) are also debiased from the projection effect i.e. they are already divided by $\sin(i)$. 

The errors are estimated from the scatter in line-free regions. We fit for the spectrum and the differential visibilities simultaneously; however, because the former is sensitive to telluric correction and has very small statistical errorbars, we increase the flux errorbars by a factor of two. We find that this scaling led to a comparable reduced $\chi^2$ between flux and visibilities in all observations. Moreover, because of the blending with the Br$\gamma$ stationary emission line, which also produces differential visibility signatures, it is necessary to perform simultaneous fits for the Br$\gamma$ line and the jets. For the model and results for the Br$\gamma$ stationary line we refer to the companion paper on the equatorial outflows (Waisberg et al., sub.). The results presented here correspond to the "outflow" model in the companion paper (which we favor over the "disk" model). 

The differential visibilities are computed with respect to the best fit continuum model (see companion paper Waisberg et al., sub.). The 2D Fourier transform of the jet model detailed above is computed semi-analytically. The total differential visibility at a given spectral channel is then 

\begin{equation}
V_{diff} (\vect{u}) = \frac{V(\vect{u})}{V_c(\vect{u})} = \frac{1 + \sum \limits_{i} \frac{V_i(\vect{u})}{V_c(\vect{u})} f_i}{1 +\sum \limits_{i}  f_i}
\end{equation}

\noindent where $V_c$ is the continuum visibility, and $f_i$ and $V_i$ are the flux ratios relative to the continuum and visibilities for each component $i$ (different jet knots, Br$\gamma$ stationary line). 

\hspace{0.5cm} The fits are done through non-linear least squares minimization with the Levenberg-Marquardt method through the \texttt{python} package LMFIT \footnote{https://lmfit.github.io/lmfit-py/}. The quoted errors correspond to the 1-$\sigma$ errors from the least squares fit i.e. the estimated derivatives around the optimal solution (scaled by $\sqrt{\chi^2_{red}}$). We caution, however, that true uncertainties are dominated by (i) degeneracies between the many parameters, which create a complicated multi-dimensional $\chi^2$ map ; (ii) systematic errors from the continuum model; (iii) the assumption of our simple "geometric" models, which cannot capture all the complexities involved. We note, in addition, that in Epoch 1 there is very severe spectral blending of the Pa$\alpha$ jet lines with the stationary Br$\gamma$ line, so that its results are less robust. 

\end{enumerate} 

\subsection*{Results}

Table \ref{table:jets} shows the model fit results for the jet emission lines in all epochs. The data and best fits are shown in Figure \ref{fig:model_plot_P99_3} for Epoch 3 of the 2017 observations and in Appendix A for all other observations. The measured position angle of the jets agrees well with that expected from the kinematic model and radio observations (Figure \ref{fig:redshift}). The results for the 2016 observation are mostly in agreement with Paper I, but the more general model emission profile allows to better constrain $r_0 = 0.16 \pm 0.02$ mas. Because of the multiple jet components, blending with stationary lines and smaller number of jet lines, the observations in 2017 are not nearly as constraining as the 2016 observation, and often only an upper limit on $r_0$ can be estimated. 

\begin{table*}[ht]
\centering
\caption{GRAVITY Jets Model Fit Results}
\label{table:jets}
\begin{tabular}{ccccccc} 
\hline \hline \\ 
Parameter & unit & Jet Knot &  \shortstack{2016-07-17} & \shortstack{2017-07-09\\Epoch 1} & \shortstack{2017-07-09\\Epoch 2} & \shortstack{2017-07-10\\Epoch 3} \\[0.3cm] 
\hline \\ 

$i_{red}$ & deg & \shortstack{Compact\\Extended} & \shortstack{$84.324 \pm 0.006$\\\vspace{2.4mm}} & \shortstack{$63.67 \pm 0.01$\\$64.88 \pm 0.01$} & \shortstack{$61.0 \pm 0.1$\\$62.09 \pm 0.03$} & \shortstack{$62.30 \pm 0.01$\\$61.21 \pm 0.02$}  \\ [0.3cm] 

$i_{blue}$ & deg &  \shortstack{Compact\\Extended} & \shortstack{$85.201 \pm 0.006$\\\vspace{2.4mm}} & \shortstack{$63.11 \pm 0.01$\\\vspace{2.4mm}} & \shortstack{$59.78 \pm 0.01$\\\vspace{2.4mm}} & \shortstack{$60.64 \pm 0.02$\\$59.85 \pm 0.02$} \\ [0.3cm] 

FWHM & km/s & All & $1230 \pm 13$ & $1700 \pm 13$ & $1661 \pm 21$ & $1705 \pm 20$ \\ [0.3cm] 

PA & deg & All & $77.2 \pm 1.5$ & $87.3 \pm 0.6$ & $87.1 \pm 1.4$ & $84.9 \pm 1.0$  \\ [0.3cm]

$\theta$ & mas &  \shortstack{Compact\\Extended} & \shortstack{$3.7 \pm 0.5$\\\vspace{2.4mm}} &  \shortstack{$1.2 \pm 0.2$\\$1.9 \pm 0.1$} & \shortstack{$3.7 \pm 0.3$\\$1.8 \pm 0.3$} &  \shortstack{$0.7 \pm 0.1$\\$19.9 \pm 17.4$}  \\ [0.3cm] 

$\alpha$ & mas &  \shortstack{Compact\\Extended} & \shortstack{$0.2 \pm 0.1$\\\vspace{2.4mm}} & \shortstack{$1.0 \pm 0.3$\\$3.6 \pm 0.2$} & \shortstack{$0.7 \pm 0.1$\\$4.8 \pm 0.5$} & \shortstack{$1.9 \pm 0.4$\\$0.1 \pm 0.4$} \\ [0.3cm] 

$r_0$ & mas &  \shortstack{Compact\\Extended} & \shortstack{$0.16 \pm 0.02$\\\vspace{2.4mm}} & \shortstack{$0.5 \pm 0.1$\\$\lesssim 3$} & \shortstack{$\lesssim 0.5$\\$\lesssim 8$} & \shortstack{$\lesssim 0.5$\\$3.1 \pm 0.1$}   \\ [0.3cm] 

$\mu$\tablefoottext{a} & mas &  \shortstack{Compact\\Extended} & \shortstack{$1.7 \pm 0.4$\\\vspace{2.4mm}} & \shortstack{$1.7 \pm 0.3$\\$\approx 6-8$} & \shortstack{$\approx 2-4$\\$\approx 6.5-13$} & \shortstack{$1.5 \pm 0.5$\\$12 \pm 8$}  \\ [0.3cm] 

\hline \\ 

$\dfrac{\chi^2}{\text{dof}}$ & - & - & 1.4 & 2.7 & 1.6 & 1.0 \\ [0.3cm]

\hline

\end{tabular}
\tablefoot{
\tablefoottext{a}{Centroid of the jet emission profile.}
}

\end{table*}

\begin{figure}[tb]
\centering
\includegraphics[width=\columnwidth]{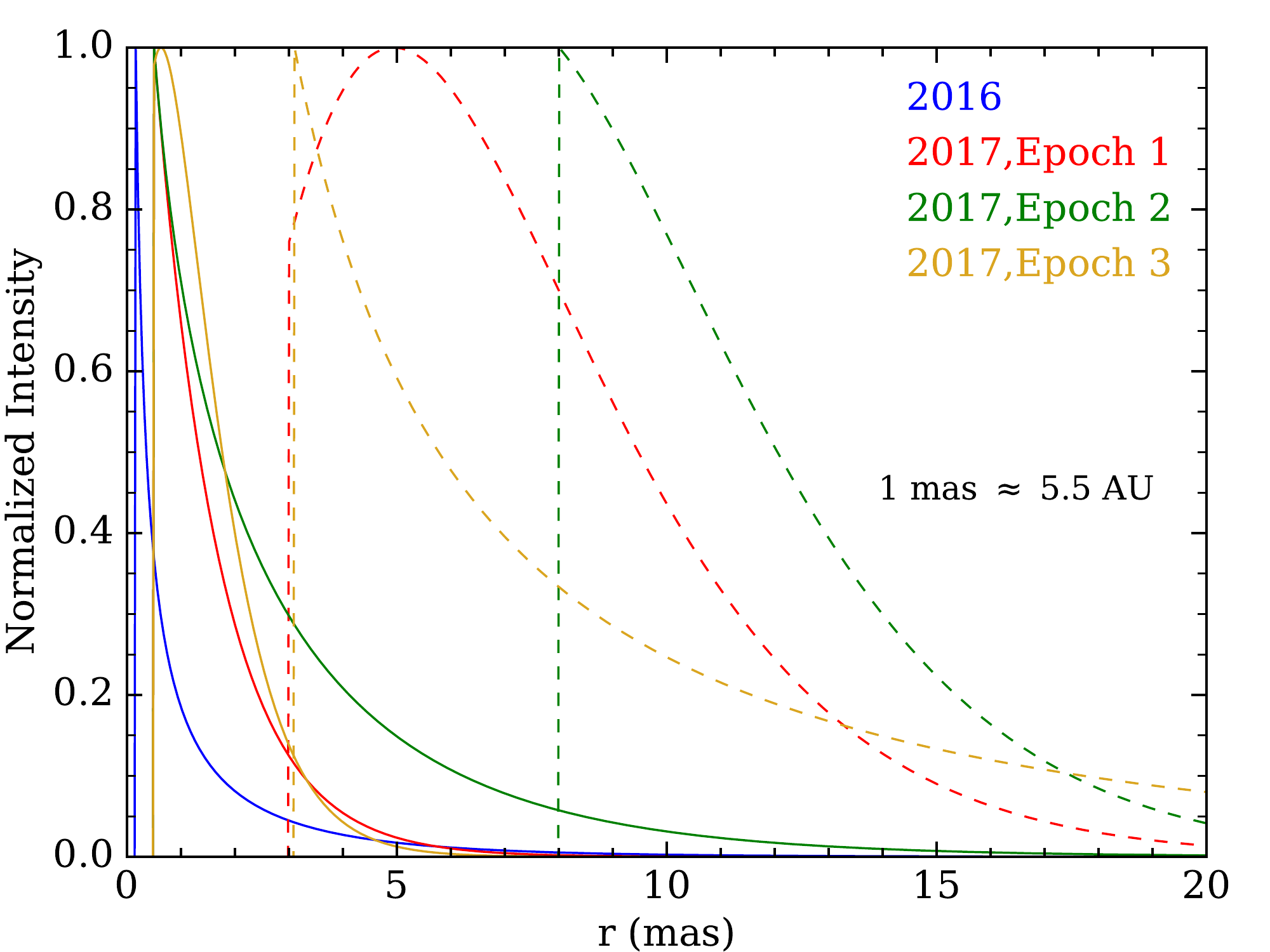} \\ 
\includegraphics[width=\columnwidth]{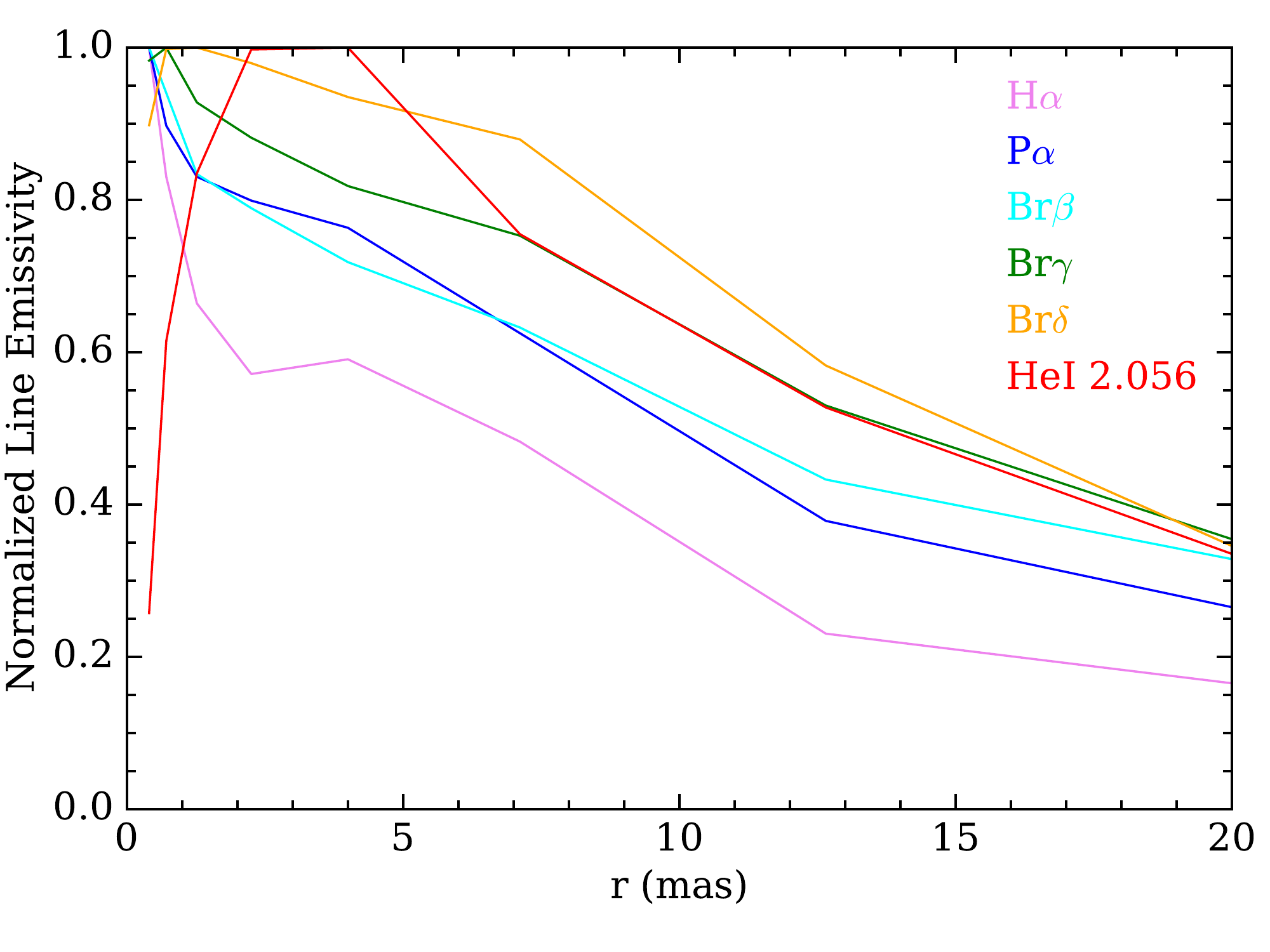}
\caption{\textbf{Top:} Collection of optical jet radial emission profiles for all GRAVITY observations. These were derived from the available jet emission lines in the K band spectrum (Br$\gamma$, Br$\delta$ and He I 2.06 $\mu$m in the 2016 observation, and Pa$\alpha$ and Br$\beta$ in the 2017 observations). There are two classes of profiles: compact (solid lines), associated with recent or current knots and which peak close to the central binary, and extended (dashed lines), associated with older knots. \newline \textbf{Bottom:} Emission profiles for selected emission lines calculated from a \texttt{Cloudy} photoionization model using the best fit parameters to the emission line fluxes from Epoch X1 of the XSHOOTER observations. The elongated profiles resolved by GRAVITY suggest photoionization by the collimated radiation as the heating mechanism. The steeper profiles relative to the model are very likely due to screening effects (large area filling factor by the bullets in the jets).}
\label{fig:jet_profile}
\end{figure}

As noted before, there are two components ("knots") in the jet lines in the 2017 observations. The need for two components in 2017 is clear from the much broader lines, as well as from the interferometric data (differential visibility amplitudes and phases which are not aligned). Two components are clearly distinguishable in all epochs for  the Pa$\alpha$ line from the receding jet but only on Epoch 3 for the Br$\beta$ line from the approaching jet. One of the components is compact (emission centroid $\lesssim 2$ mas), with an emission profile similar to the one in the 2016 observation, and associated with the most recent or current jet ejection. The other component is significantly more extended (emission centroid $\gtrsim 6$ mas), associated with a previous jet ejection. This interpretation agrees with the corresponding redshifts from the precession model in Epochs 1 and 2. Figure \ref{fig:jet_profile} shows the collection of radial emission profiles of the optical jets for all observations. The compact knots have an exponential-like profile, whereas the extended knots have more rotund shapes. 

We find that for the more compact jet components, the emission peaks substantially close to the binary system, and is more compact than previous estimates from optical spectroscopy monitoring, e.g. \cite{Borisov87} derived an exponential decay of the jet emission with falloff distance $6.7 \times 10^{14} \text{ cm} \approx 8.2$ mas for the H$\alpha$ jet line. The elongated emission profiles that we measure are strongly suggestive of a continuous heating mechanism along the entire jet, which is naturally accomplished by photoionization by the collimated radiation from the inner regions close to the compact object. As in Paper I, we attempted to fit the jets with more localized emission profiles (such as a point source or Gaussian), but they are inconsistent with the data: an elongated structure for the jets is strongly preferred. 

Furthermore, the 2017 observations allow to probe the spatial evolution of the jet profiles from night to night. The jet lines in Epoch 1 have clearly disappeared by Epoch 2 two nights later (Figure \ref{fig:spectrum}). However, the jets lines in Epochs 2 and 3 (which are separated by one night only) partly overlap. The extended component in Epoch 3 could correspond to the compact component in Epoch 2 after it has travelled $\approx 8 \text{ mas/day } \times 1 \text{ day } \times \sin(60^\circ) \approx 7 \text{ mas}$ on the sky plane, whereas a new compact component in Epoch 3 clearly appeared at the same redshift where the extended component in Epoch 2 was located. This is the first time the movement of the optical bullets has been spatially resolved, although longer observations during a single night would be needed to trace the motion of an individual component unambiguously. 

\section{XSHOOTER DATA ANALYSIS}
\label{Xshooter} 

We fit the jet lines with Gaussian or Lorentzian profiles (the jet lines are usually better fit by a Gaussian, but sometimes a Lorentzian profile is clearly preferred) to estimate their central wavelength, FWHM and total intensity. All line fits were performed with the task \texttt{splot} in \texttt{IRAF}\footnote{IRAF is distributed by the National Optical Astronomy Observatories, which are operated by the Association of Universities for Research in Astronomy, Inc., under cooperative agreement with the National Science Foundation.} \citep{Tody86,Tody93}. Whenever jet lines were blended with other jet or stationary lines, deblending was used. For all line identifications and adopted wavelengths, the \href{http://www.pa.uky.edu/~peter/newpage/}{"Atomic Line List v2.05b21"} (van Hoof, P.) was used. The noise for the spectral fits was estimated from the scatter in continuum regions near each line. A particularly important constraint for photoionization models of the optical bullets is the absence of ionized helium in the jets due to the lack of the He II 4686$\AA$ line. We estimate upper limits on its flux from the known location of where the line would appear based on the measured redshifts and the FWHM from the other jet lines. 

\begin{figure*}[tb]
\centering
\includegraphics[width=2.0\columnwidth]{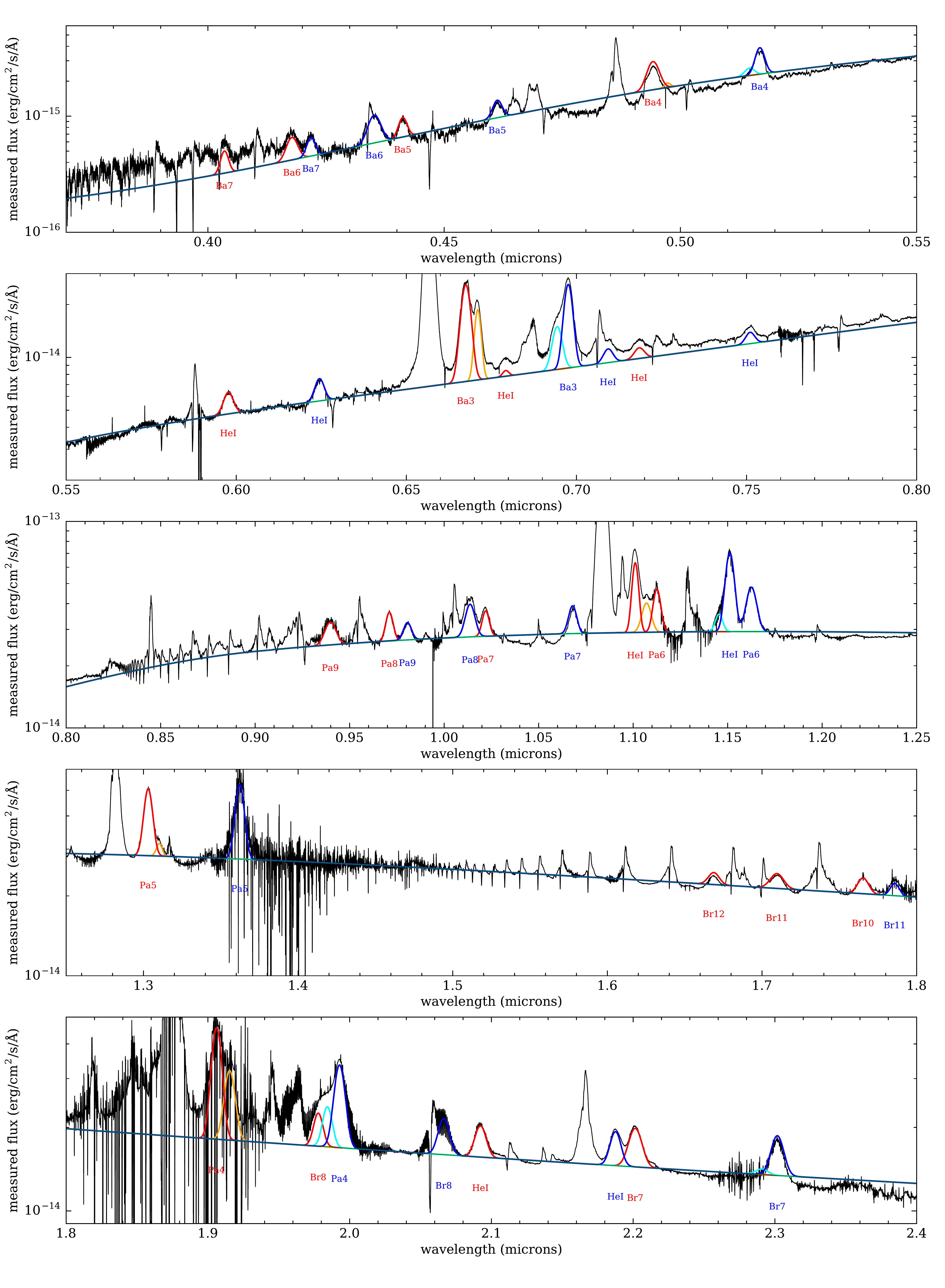} \\ 
\caption{First XSHOOTER spectrum of SS 433 (Epoch X1). H I and He I emission lines from the eastern and western optical jets are shown in blue and red, respectively. Older jet ejections (cyan and orange), only present for some line transitions such as H$\alpha$ and Pa$\alpha$, are also shown.}
\label{fig:spectrum_XSHOOTER}
\end{figure*}

Figure \ref{fig:spectrum_XSHOOTER} shows the spectrum for the first XSHOOTER observation of SS 433 (Epoch X1), which contains the largest number of jet lines of all epochs. Emission lines from the jet are shown in blue and red for the eastern and western jets, respectively (at this epoch, the eastern jet, which is approaching most of the time, is receding). Throughout all observations we identify up to twenty pairs of lines: H$\alpha$ through H$\epsilon$; Pa$\alpha$ through Pa9; Br$\beta$ through Br12; and five lines of He I (5875$\AA$, 6678$\AA$, 7065$\AA$, 1.083 $\mu$m, 2.056$\mu$m). The strongest hydrogen lines (H$\alpha$, H$\beta$, Pa$\alpha$, Br$\beta$) and He I 1.083$\mu$m often show one or two additional components (shown in cyan and orange in Figure \ref{fig:spectrum_XSHOOTER}), with redshift corresponding to previous ejections according to the precession model (and which we associate with older jet knots), whereas most emission lines show only one component from the most recent or current ejection (this behavior matches what is seen in the GRAVITY observations). The average redshift for the more recent jet knots in each epoch is shown in Figure \ref{fig:redshift} and agrees with the kinematic precession model within the random jitter. The strength and number of jet emission lines varies significantly between the five XSHOOTER epochs. 

\subsection*{\texttt{Cloudy} Models for Photoionized Optical Bullets}

The spatial emission profiles for the optical jets resolved by GRAVITY are strongly suggestive of photoionization as the main heating mechanism of the optical jets. Therefore, we proceeded to fit the jet emission line fluxes measured from XSHOOTER with photoionization models.

All line emission in SS 433 originates from dense gas; a clear lower limit to the density $n_e \gtrsim 10^7 \text{cm}^{-3}$ follows from the absence of any forbidden lines in the spectrum \citep[][p.60]{AGN3}. The true density in the optical bullets is estimated to be $\sim 10^{13} \text{cm}^{-3}$, and the optical jets take the form of dense bullets distributed throughout the jets with a low volume filling factor $\sim 10^{-6}$ \citep{Fabrika04}. At such densities, collisional ionization from excited levels and radiative transfer effects become important, causing strong deviations from Case B recombination; still, such densities are not high enough for LTE to apply to all levels, and full collisional radiative models must be used \citep{Ferland13,Ferland17}. A clear sign that collisional and radiative transfer effects are important in the optical bullets are the rather flat or even inverted H$\alpha$/H$\beta$ jet line ratios (after extinction correction). 

We model the jet emission lines with the photoionization code \texttt{Cloudy} \citep{Ferland13,Ferland17} version c17.01. The ionizing radiation is taken to be a single blackbody. We construct a grid in bullet density, bullet size, blackbody temperature and blackbody intensity. In addition, because SS 433 is known to have appreciable extinction, we expand the grid in $A_V$, assuming an extinction law $\frac{A_{\lambda}}{A(V)}$ following \cite{Cardelli89} and $R_V = 3.1$. We attempted to fit for the He abundance, since it is likely that the donor star in SS 433 is at an advanced evolutionary stage, but it was unconstrained in our fits; therefore, we fixed a standard He abundance ($0.098$ by number). The grid parameters and intervals are shown in Table \ref{table:Cloudy}. The solutions are iterated until convergence of the optical depths. Each point in the grid gives a model intensity for each line (erg/s/$\text{cm}^2$). The total number of bullets is then computed to minimize the $\chi^2$ with respect to the measured line intensities. The bullets are assumed to be identical in both jets, but the number of bullets (and therefore the kinetic power) can be different for the receding and approaching jets. We corrected the line intensities in each jet for Doppler boosting $(1+z)^3$ using the average measured redshift. 

\begin{table*}[t]
\centering
\caption{\texttt{Cloudy} Model Grid Parameters}
\label{table:Cloudy} 
\begin{tabular}{cccccc}
\hline \hline 
Parameter & Symbol & Unit & Minimum & Maximum & Interval \\[0.3cm]
\shortstack{Blackbody\\Temperature} & $T$ & $10^4$ K & 1 & $10^2$ & 0.1 (Log) \\ [0.3cm]
\shortstack{Blackbody\\Intensity} & $I$ & $10^{13}$ erg/s/$\text{cm}^2$ & $10^{-5}$ & $10^4$ & 0.5 (Log) \\ [0.3cm]
\shortstack{Bullet\\Density} & $n_H$ & $10^{13}$ $\text{cm}^{-3}$ & $10^{-3}$ & $10^2$ & 0.5 (Log) \\ [0.3cm]
\shortstack{Bullet\\Size} & $R$ & $10^6$ cm & $10^{-2}$ & $10^2$ & 0.5 (Log) \\ [0.3cm]
Extinction & $A_V$ & - & 5 & 8 & 0.2 \\ [0.3cm]
\hline
\end{tabular}
\end{table*}

We use the lines from the most recent jet knots (which contain a much larger number of lines) for the fits. The GRAVITY observations revealed that recent ejections have a typical exponential intensity profile as shown in Figure \ref{fig:jet_profile}. We integrate our \texttt{Cloudy} models over five radial points covering a factor of $100$ in intensity. The reported best fit intensity is the one at the base of the optical jet. 

The statistical line flux errors estimated from the spectral fits are very small. The actual errors are dominated by systematic effects in the flux calibration, telluric correction, contamination by weak lines, line blending and model uncertainties (such as collisional cross sections). We estimate a $25\%$ error in flux for all the lines, which results in $\chi_{red}^2 \sim 1$ for the best fit grid points. The parameter uncertainties are estimated by considering all the models which fall within $\Delta \chi^2 = p$ of the best $\chi^2$ \citep{Lampton76}, where $p$ is the number of model parameters. 

From the \texttt{Cloudy} model results, we derive further parameters of interest for the jets: 

\begin{enumerate}

\item The volume filling factor of the optical bullets: 

\begin{equation}
V_{\rm ff} = \frac{V_{\rm bullets}}{V_{\rm jet}}
\end{equation}

\noindent where $V_{\rm bullets}$ is the total volume in the bullets (calculated from their size and number) and $V_{\rm jet} = \pi \psi_{\rm jet}^2 \dfrac{l_{\rm jet}^3}{3}$, where $\psi_{\rm jet}$ is the half-opening angle and $l_{\rm jet}$ is the length of the jet;  

\item The kinetic power of the optical jets:

\begin{equation}
L_{\rm kin} = 2 \times \frac{1}{2} M v_{\rm jet}^2 \frac{v_{\rm jet}}{l_{\rm jet}}
\end{equation}

\noindent where $v_{\rm jet} = 0.26c$, $M$ is the total mass in the bullets (calculated from their density, size and number). 

\item The total line luminosity in the optical bullets $L_{\rm bullets}$. This includes not only the recombination lines within XSHOOTER, but all of the strongest hydrogen and helium lines in the full spectrum output from \texttt{Cloudy}. 

\item The total ionizing luminosity within the beam containing the optical bullets:  

\begin{equation}
L_{\rm beam} = 2 \times I (r_0) \pi \psi_{\rm jet}^2 r_0^2 
\end{equation} 

\noindent where $I(r_0)$ is the intensity at the base of the beam. The total luminosity in the collimated radiation could be higher if it is broader than the beam containing optical bullets, which is likely the case from the presence of older jet knots that can keep radiating for a few days \citep{Panferov93}.  

\item The luminosity inferred by an observer (assuming isotropy) whose line of sight is within the collimated beam: 

\begin{equation}
L_{\rm face-on} = L_{\rm beam} \frac{4 \pi}{\pi (2 \psi_{\rm jet})^2} = 2 I (r_0) r_0^2 
\end{equation}

\end{enumerate} 

For these estimates, we assumed $\psi_{\rm jet} = 1^\circ$ and $r_0 = 0.4$ mas and $l_{\rm jet} = 2$ mas from the GRAVITY observations. 

Our model does not take into account screening of radiation by the bullets, which plays a role because the jets are very compact (area filling factors are high). To ensure self-consistency in the energetics, we only accept solutions with enough total luminosity in the beam to power the bullets (i.e. $L_{\rm beam} > L_{\rm bullets}$). We also note that we assume the bullet radiation is isotropic and identical between the two jets (the only difference we allow between the two jets is in the total number of bullets), whereas differences between the two jets with precessional phase have been interpreted in terms of anisotropic radiation \citep{Panferov97b,Fabrika04}.

\subsection*{Results}

We note that several related calculations to estimate the properties of the optical jets can be found in previous papers \citep[e.g.][]{Davidson80,Begelman80,Bodo85,Borisov87,Fabrika87,Brown91,Panferov93}, primarily based on the H$\alpha$ luminosity. These calculations have several uncertainties: unknown heating mechanism/line emissivity, degeneracy between density and volume filling factor, unknown extinction, optical depth effects, unknown spatial emission profile of the optical jets (e.g. jet size). The calculations presented here are based on the first optical interferometric measurements that have spatially resolved the optical jets, and on a large number of jet line species from XSHOOTER spectra. 

\begin{figure*}[tb]
\centering
\includegraphics[width=1.8\columnwidth]{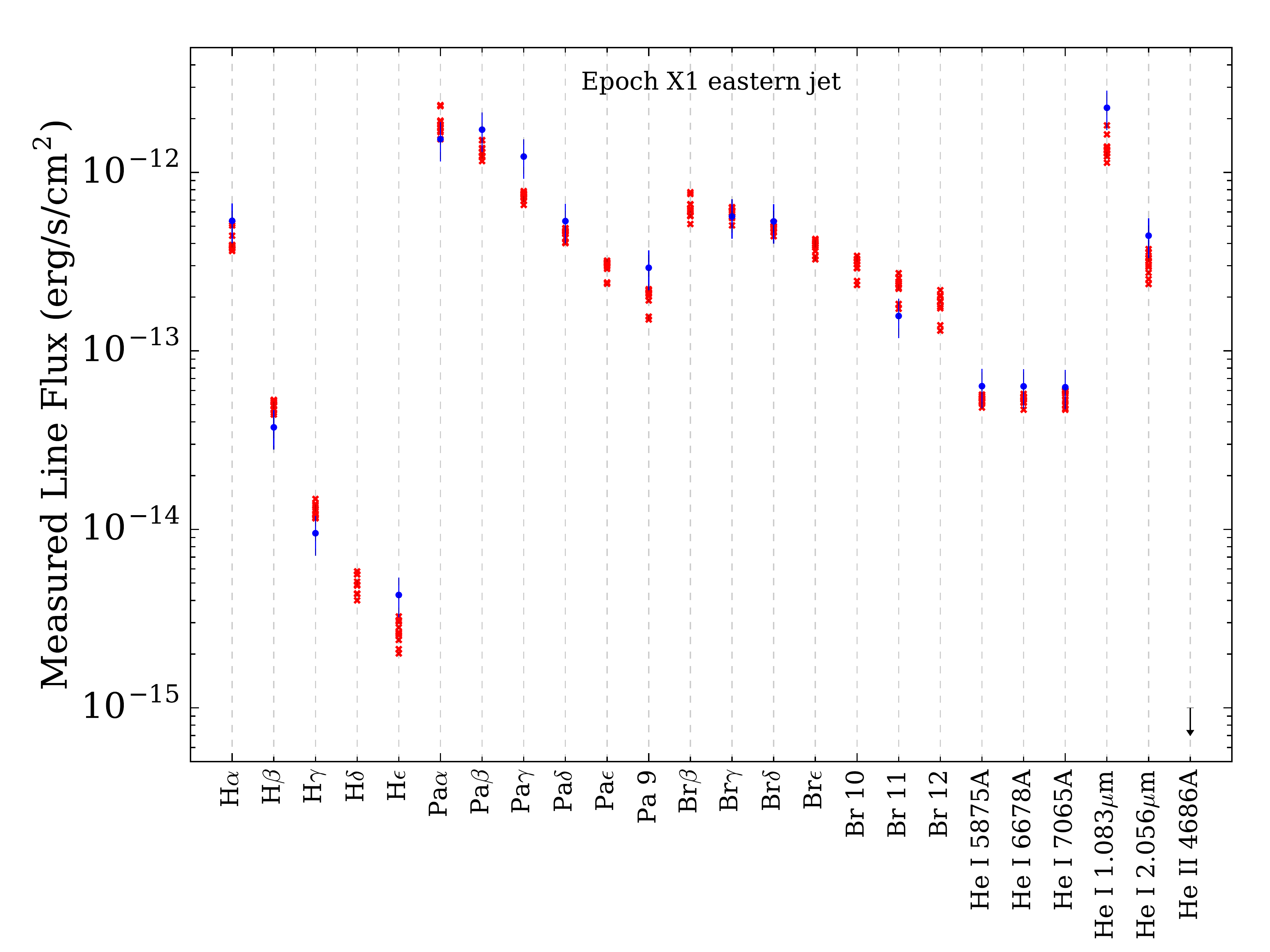}\\
\includegraphics[width=1.8\columnwidth]{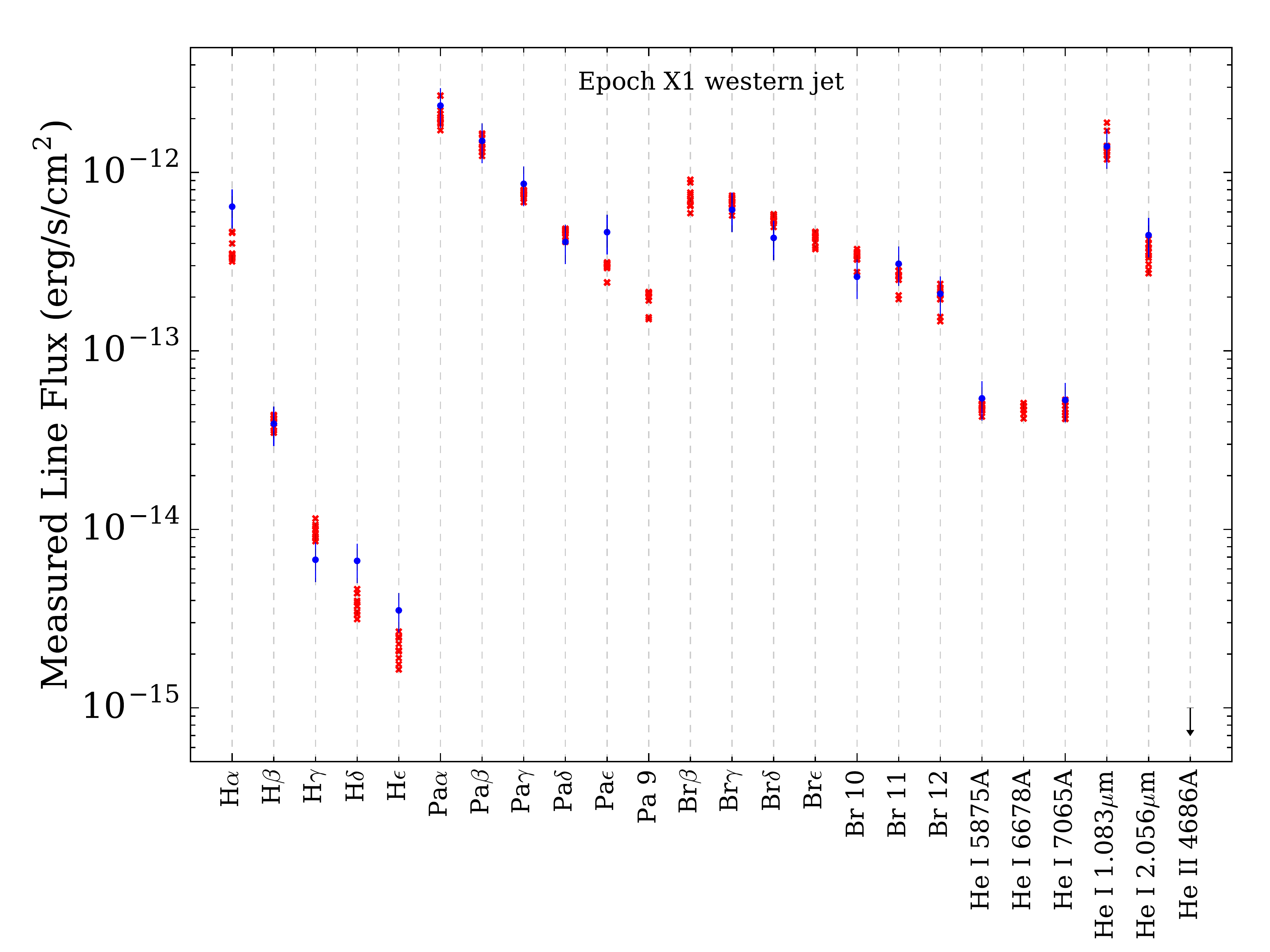}
\caption{Measured line fluxes for the Epoch X1 XSHOOTER observation of SS 433 (blue) and  best fit \texttt{Cloudy} photoionization models (red). The models are shown for all the lines measured in at least one of the five epochs. The He II $4686 \AA$ line flux is an upper limit since it was never detected.}
\label{fig:model_XSHOOTER_1}
\end{figure*}

Table \ref{table:Cloudy_results} shows the model fit results for the five XSHOOTER epochs. Figure \ref{fig:model_XSHOOTER_1} shows the measured and model line fluxes for the first XSHOOTER epoch (corresponding plots for the other epochs are shown in Appendix B). 

\begin{table*}[t]
\centering
\caption{\texttt{Cloudy} Photoionization Models for the Optical Jets: Results}
\label{table:Cloudy_results} 
\begin{tabular}{ccccccc}
\hline \hline \\
Parameter & Unit & Epoch X1 & Epoch X2 & Epoch X3 & Epoch X4 & Epoch X5 \\ [0.3cm] 
\hline \\ 

\shortstack{Number of\\Jet Lines Fitted\\(Eastern Jet/ Western Jet)} & - & $19/17$ & $12/13$ & $4/7$ & $9/9$ & $7/12$   \\[0.3cm]

\multicolumn{5}{c}{Model Parameters} \\[0.3cm]

$T$ & $10^4$ K & $3-4$ & $3-4$ & $3-4$ & $3$ & $3-5$ \\ [0.3cm]

$I(r_0)$ & $10^{13}$ erg/s/$\text{cm}^2$ & $1-3$ & $1$ & $1-3$ & $1-3$ & $1-3$ \\ [0.3cm]

$n_H$ & $10^{13} \text{ cm}^{-3}$ & $1-3$ & $0.3-3$ & $0.3-10$ & $1$ & $0.3-3$ \\ [0.3cm]

$R$ & $10^6$ cm & $1-30$ & $1-100$ & $1-100$ & $3$ & $1-100$ \\ [0.3cm]

$A_V$ & - & $6.4-6.8$ & $6.2-6.8$ & $6.6-7.6$ & $6.6-7.4$ & $6.6-7.6$ \\ [0.3cm]

\shortstack{$N_{\rm bullets}$\\(both jets)} & $10^{13}$ & $0.1-300$ & $0.02-20$ & $0.02-40$ & $70-100$ & $0.05-450$ \\ [0.3cm]

$\dfrac{\chi^2}{\text{dof}}$ & - & $42/29$ & $44/18$ & $8/4$ & $36/11$ & $35/12$ \\ [0.3cm]

\hline \\ 
\multicolumn{5}{c}{Derived Parameters} \\[0.3cm]

\shortstack{$V_{ff}$\\(average both jets)} & $10^{-6}$ & $1-20$ & $0.6-100$ & $0.5-160$ & $8-12$ & $1-160$  \\ [0.3cm]

\shortstack{$L_{\rm kin}$\\(both jets)} & $10^{38}$ erg/s & $2.5-14$ & $1.4-24$ & $2-47$ & $5-9$ & $3-35$  \\ [0.3cm]

\shortstack{$L_{\rm bullets}$\\(both jets)} & $10^{37}$ erg/s & $1.2-1.9$ & $0.8-1.7$ & $1.2-4.0$ & $1.6-3.6$ & $1.6-3.2$ \\ [0.3cm]

\shortstack{$L_{\rm beam}$\\(both jets)} & $10^{37}$ erg/s & $2-6$ & $2$ & $2-6$ & $2-6$ & $2-6$  \\ [0.3cm]

\shortstack{$L_{\rm face-on}$\\(both jets)} & $10^{40}$ erg/s & $7-20$ & $7$ & $7-20$ & $7-20$ & $7-20$ \\ [0.3cm]

\hline 
\end{tabular}
\end{table*}

\subsubsection*{The Bullet Properties}

The results confirm that the optical bullets in SS 433 are very dense $n_H \sim 10^{13} \text{cm}^{-3}$ and have a size $R \approx 10^6 - 10^7 \text{cm}$. 
Even though they are optically thin to electron scattering ($\tau_T \sim 0.01$), many emission lines are optically thick. For the best fit model in Epoch X1, $\tau_{\rm H\alpha} \approx \tau_{\rm P\alpha} \approx 20$, $\tau_{\rm H\beta} \approx 4$, $\tau_{\rm Br\gamma} \approx 2$, $\tau_{\rm Br10} \approx 0.3$, $\tau_{\rm HeI 6578} \approx 5$ and $\tau_{\rm HeI 1.083} \approx 0.6$, so that optical depth effects in the line emission indeed have to be taken into account. The cooling in the bullets is dominated by hydrogen and helium recombination lines ($\sim 80\%$), with total line luminosity $\approx 2 \times 10^{37}$ erg/s. Only around $15\%$ of such luminosity is in lines within the XSHOOTER spectrum, with the strongest lines being the Lyman series in the UV. 

\subsubsection*{The Jet Kinetic Power}

We constrain the kinetic power in the optical bullets to $\sim 2-20 \times 10^{38}$ erg/s. This agrees with estimates from emission line modeling in the X-rays, 
which vary from from $\sim 3 \times 10^{38}$ to $5 \times 10^{39}$ erg/s \citep[e.g.][]{Marshall02,Brinkmann05,Medvedev10}. The collapse of the continuous X-ray jets into optical bullets must therefore be an efficient process, both in terms of mass and kinetic energy. Previous estimates from optical spectroscopy give $L_{\rm kin} \sim 10^{39}$ erg/s \citep[e.g.][]{Panferov97a,Fabrika87}.

\subsubsection*{The Collimated Radiation}

The results constrain the collimated radiation to be relatively soft $T \sim 3-4 \times 10^4$ K and the total luminosity in the $1^\circ$ beam containing the optical bullets to $L_{\rm beam} \approx 2-6 \times 10^{37}$ erg/s. Lower luminosities cannot power the jet emission lines, whereas higher luminosities and temperatures cause too intense heating and ionize helium too much. For an observer looking face on at the collimated radiation and assuming isotropy, the inferred luminosity would be $\approx 7-20 \times 10^{40}$ erg/s i.e. SS 433 would appear as an \textit{extremely} bright UV source. This luminosity is higher than the $\sim \text{few} \times 10^{39}$ erg/s inferred from the SED \citep[][although this is a rather uncertain number -- see below]{Wagner86}, suggesting that indeed there is collimated radiation in SS 433 (not only thermal downgrading at low latitudes). The total luminosity in collimated radiation depends on its opening angle, and would be $\sim 2 \times 10^{39} - 10^{41}$ erg/s for angles $10^\circ - 50^\circ$. From modeling of the optical filaments in the W50 nebula and assuming photoionization by collimated radiation, \cite{Fabrika08} estimated an ionizing luminosity $\sim 10^{40}$ erg/s in an opening angle $\sim 50^\circ$. 

Figure \ref{fig:jet_profile} (bottom) shows the spatial emission profile for selected emission lines for the best fit \texttt{Cloudy} photoionization model to Epoch X1. It assumes the bullets are distributed homogeneously over the jet with constant properties (density, size), and shows the normalized line emissivity as a function of distance $r$ in the jet. It resembles the spatial profiles directly resolved by GRAVITY, and confirms that the different H I and He I lines have similar emission regions. The elongated profile results from the interplay between the radiation intensity and gas density, and is a good consistency check for our photoionization models. A caveat is that we do not consider screening of the intensity along the jet by the bullets (large area filling factors by the bullets in the jet), which is likely the reason for the steeper measured profiles relative to the model.

\begin{figure}[tb]
\centering
\includegraphics[width=\columnwidth]{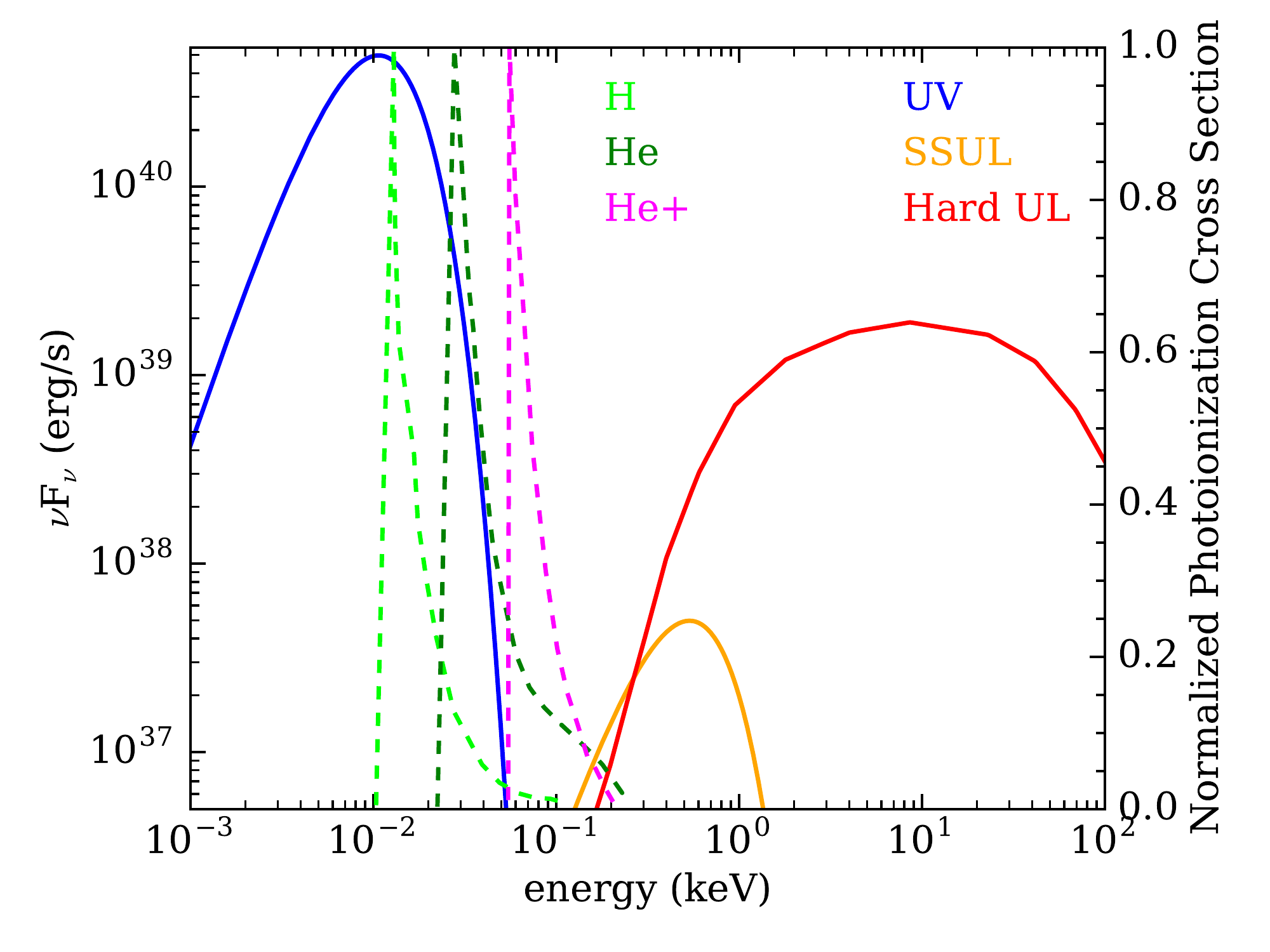}\\
\caption{Potential SS 433 SED for a face on observer who sees the beam radiation as the optical bullets see it. The solid blue curve shows the UV component from the best fit \texttt{Cloudy} photoionization model to the line intensities in Epoch X1 of XSHOOTER. The orange and red lines show the upper limit to a possible X-ray component for a Supersoft Ultraluminous (SSUL) and Hard Ultraluminous (Hard UL) type SEDs, respectively. The dashed lines show the ground state photoionization cross sections for H, He and He+. For SS 433 to look like an ULX for the face on observer, its X-ray spectrum must be significantly hard, since soft X-ray photons break the H/He/He+ ionization balance in the optical bullets.}
\label{fig:ULX}
\end{figure}

Because of the possibility that SS 433 could be an ULX for an observer looking at it face on, we attempted to constrain the X-ray luminosity in the beam containing the optical bullets. We did this through a perturbative approach: we selected the best fit model in Epoch X1 from the analysis above, and added a second X-ray component with varying intensity, and checked at which X-ray intensity the optical emission lines are substantially affected and in clear violation of the data (e.g. by producing a too strong He II 4868 $\AA$ line). Because the main effect comes from the soft X-ray photons capable of ionizing H, He and He+, the constraint depends on the relative contribution of soft X-ray photons to the total X-ray luminosity. To this end, we repeated the procedure for two different SEDs: a very soft one corresponding to a Supersoft Ultraluminous (SSUL) source, in the form of a blackbody $kT = 0.14$ keV, and a harder one corresponding to a Hard Ultraluminous (HL) source, in the form of a blackbody $kT= 0.27$ keV plus a strong hard component \citep{Kaaret17}. The corresponding SEDs and limits are shown in Figure \ref{fig:ULX}. 

In the case of the SSUL SED, we constrain the X-ray luminosity to be $\lesssim 10^{-3}$ of the UV component i.e. $\lesssim 5 \times 10^{34}$ erg/s in the $1^\circ$ beam containing the optical bullets or $\lesssim 10^{38}$ erg/s for a face on observer. In this case, SS 433 would not be an ULX. For the hard UL SED, the X-ray luminosity could be much larger, up to $\lesssim 10^{-1}$ of the UV component i.e. $\lesssim 5 \times 10^{36}$ erg/s in the $1^\circ$ beam containing the optical bullets or $\lesssim 10^{40}$ erg/s for a face on observer. We conclude that SS 433 could be an ULX, as long as its X-ray spectrum is dominated by hard X-rays. Just the same, face on ULXs (with a clear view through the funnel) are generally expected to have dominantly X-ray spectra \citep{Kaaret17}, whereas our results suggest that SS 433 is UV-dominated even in the collimated beam. This might mean that thermal downgrading happens already in the funnel of SS 433 \citep{Begelman06}. The most promising way to find face on SS 433-like objects in other galaxies might be to look for very bright and variable (due to jet precession) UV sources. 

The X-ray luminosity of SS 433 has also been constrained from a putative reflection component $\sim 10^{35}$ erg/s in the hard X-ray spectrum $\gtrsim 10$ keV of SS 433 \citep{Medvedev10,Middleton18}. \cite{Middleton18} estimates that the intrinsic X-ray luminosity is $L_X \gtrsim 10^{38}$ erg/s. This is consistent with our upper limits as long as the X-ray radiation is slightly less collimated ($\gtrsim 5^\circ$) than the beam containing the optical bullets, which is almost certainly the case from the presence of older jet knots that keep radiating for a few days \citep{Panferov93}. 

\subsubsection*{The Extinction Towards SS 433} 

The extinction towards SS 433 is known to be large from the very red continuum spectrum but its exact value is difficult to assess. 
Galactic dust extinction maps give $A_V = 7.8$ towards the direction of SS 433 \citep{Schlegel98,Perez10}, but that is an upper limit to 
the total integrated line of sight extinction. More recent 3D dust maps rather give $A_V = 5.7 \pm 0.1$ at $d = 5.5 \text{ kpc}$ towards SS 433 \citep{Green18}. Strong Diffuse Interstellar Bands (DIBs) are also present in the spectrum of SS 433 \citep[as noted in previous work e.g.][]{Murdin80,Margon84}. The strength of DIBs has been shown to be correlated with interstellar extinction \citep[e.g.][]{Herbig75,Friedman11,Kos13}, although with substantial scatter. We measured the EWs of several DIBs (5780, 5797, 5850, 6196, 6202, 6270, 6283, 6379, 6613, 6660 $\AA$) in our XSHOOTER spectra and used the correlations in \cite{Friedman11} and \cite{Kos13} to estimate $A_V = 5.1 \pm 1.0$ and $A_V = 5.6 \pm 2.1$, respectively, where the uncertainties are the $1\sigma$ scatter between the different DIBs. Therefore, there is evidence that the interstellar extinction towards SS 433 may be $A_V \lesssim 6.0$. 

On the other hand, $A_V \gtrsim 7.8-8.0$ has also been estimated from fitting the spectral energy distribution with a single reddened blackbody \citep[e.g.][]{Murdin80,Wagner86,Dolan97}. This approach, however, suffers from several problems (i) in the Rayleigh-Jeans range, the temperature of the blackbody and the extinction are very strongly correlated; (ii) there are numerous and very strong emission lines in SS 433; (iii) it is not clear whether the supercritical disk should look like a single blackbody e.g. the temperature seems to change with precession phase \citep{Wagner86}. Our XSHOOTER observations are not optimized for SED continuum fitting, but we confirm the strong degeneracy between blackbody temperature and extinction. $T \gtrsim 20000 \text{ K}$ (a reasonable expectation from the presence of He II 4868 $\AA$ stationary line) requires $A_V \gtrsim 7.5$. 

Our modeling of the jet lines yields $A_V \approx 6.7 \pm 0.1$, which is intermediate between the lower values inferred from 3D dust maps and DIBs and 
those estimated from the very reddened SED. We suggest that there may be substantial $A_V \gtrsim 1$ and structured circumstellar extinction in SS 433, 
affecting the equatorial part of the system more than the optical jets. It may be caused by dust forming from the equatorial outflows seen in radio \citep{Paragi99,Blundell01} 
and near-infrared stationary emission lines (see companion paper on equatorial outflows, Waisberg et al., sub.). Mid-infrared observations of SS 433 show evidence of dust from excess emission at $\lambda \gtrsim 20 \mu$m \citep{Fuchs06}. We speculate that mid-infrared interferometric observations with VLTI+MATISSE \citep{Lopez14} might resolve an extended dust torus in SS 433. 

\section{CONCLUSIONS} 
\label{Conclusion}

We presented a second set of GRAVITY observations of SS 433 after Paper I, as well as the first XSHOOTER observations of this object, focusing on the optical jets. We summarize our main conclusions from the GRAVITY observations as follows: 

\begin{enumerate}

\item The optical jets have elongated, exponential-like spatial emission profiles, suggestive of a continuous heating process throughout the entire jet; we argue for photoionization by collimated radiation; 

\item We have spatially resolved the movement of the optical bullets for the first time, finding more extended jet knots corresponding to previous jet ejections. 

\end{enumerate} 

\noindent Using the up to twenty simultaneous pairs of measured jet line fluxes in the XSHOOTER observations, we have constrained properties of the optical bullets and the putative ionizing radiation with \texttt{Cloudy} photoionization models: 

\begin{enumerate}

\item The optical bullets are dense $\sim 10^{13} \text{cm}^{-3}$ and have a size $\sim 10^{6} - 10^7$ cm, from which optical depth effects in the jet emission lines are important; 

\item The kinetic power of the optical jets is $\sim 2-20 \times 10^{38}$ erg/s; 

\item The beamed radiation is dominantly UV with a luminosity $\approx 2-6 \times 10^{37}$ erg/s in the $1^\circ$ beam containing the optical bullets. An observer looking directly at the beam would infer an isotropic luminosity $\approx 7-20 \times 10^{40}$ erg/s i.e. SS 433 would appear as an extremely bright UV source; 

\item In the photoionization picture, SS 433 could still be an ULX with a face on observer inferring $L_X \lesssim 10^{40}$ erg/s, as long as the X-ray SED is dominantly hard, since soft X-ray photons destroy the H/He/He+ ionization balance in the optical bullets; 

\item We constrain the extinction in the optical jets $A_V = 6.7 \pm 0.1$ and suggest there is substantial and structured circumstellar extinction in this object. 

\end{enumerate} 

\begin{acknowledgements}
We thank the GRAVITY Co-Is, the GRAVITY Consortium and ESO for developing and operating the GRAVITY instrument. In particular, I.W. and J.D. thank the MPE GRAVITY team, in particular F. Eisenhauer, R. Genzel, S.Gillessen, T. Ott, O. Pfhul and E. Sturm. We also thank the GRAVITY team members (W. Brandner, F. Einsenhauer, S. Hippler, M. Horrobin, T. Ott, T. Paumard, O. Pfhul, O. Straub, E. Wieprecht) and ESO staff who were on the mountain during the observations. We also thank P. Kervella for comments on the paper. I.W. thanks the organizers and participants of the 2018 Cloudy Workshop in Chiang Mai, Thailand, where part of this project was done, in particular Gary Ferland, Christophe Morisset and Peter van Hoof. POP  acknowledges financial support from the CNRS High Energy National Program (PNHE). POP and GD  acknowledge financial support from the CNES. This research has made use of the Jean-Marie Mariotti Center \texttt{SearchCal} service \footnote{Available at http://www.jmmc.fr/searchcal} co-developped by LAGRANGE and IPAG, CDS Astronomical Databases SIMBAD and VIZIER \footnote{Available at http://cdsweb.u-strasbg.fr/}, NASA's Astrophysics Data System Bibliographic Services, NumPy \citep{van2011numpy} and matplotlib, a Python library for publication quality graphics \citep{Hunter2007}.
\end{acknowledgements}

\bibliographystyle{aa}
\bibliography{mybib.bib}

\begin{thebibliography}{69}
\expandafter\ifx\csname natexlab\endcsname\relax\def\natexlab#1{#1}\fi

\bibitem[{{Begelman} {et~al.}(2006){Begelman}, {King}, \&
  {Pringle}}]{Begelman06}
{Begelman}, M.~C., {King}, A.~R., \& {Pringle}, J.~E. 2006, \mnras, 370, 399

\bibitem[{{Begelman} {et~al.}(1980){Begelman}, {Sarazin}, {Hatchett}, {McKee},
  \& {Arons}}]{Begelman80}
{Begelman}, M.~C., {Sarazin}, C.~L., {Hatchett}, S.~P., {McKee}, C.~F., \&
  {Arons}, J. 1980, \apj, 238, 722

\bibitem[{{Blundell} \& {Bowler}(2004)}]{Blundell04}
{Blundell}, K.~M. \& {Bowler}, M.~G. 2004, \apjl, 616, L159

\bibitem[{{Blundell} {et~al.}(2001){Blundell}, {Mioduszewski}, {Muxlow},
  {Podsiadlowski}, \& {Rupen}}]{Blundell01}
{Blundell}, K.~M., {Mioduszewski}, A.~J., {Muxlow}, T.~W.~B., {Podsiadlowski},
  P., \& {Rupen}, M.~P. 2001, \apjl, 562, L79

\bibitem[{{Bodo} {et~al.}(1985){Bodo}, {Ferrari}, {Massaglia}, \&
  {Tsinganos}}]{Bodo85}
{Bodo}, G., {Ferrari}, A., {Massaglia}, S., \& {Tsinganos}, K. 1985, \aap, 149,
  246

\bibitem[{{Borisov} \& {Fabrika}(1987)}]{Borisov87}
{Borisov}, N.~V. \& {Fabrika}, S.~N. 1987, Soviet Astronomy Letters, 13, 200

\bibitem[{{Brinkmann} {et~al.}(1988){Brinkmann}, {Fink}, {Massaglia}, {Bodo},
  \& {Ferrari}}]{Brinkmann88}
{Brinkmann}, W., {Fink}, H.~H., {Massaglia}, S., {Bodo}, G., \& {Ferrari}, A.
  1988, \aap, 196, 313

\bibitem[{{Brinkmann} \& {Kawai}(2000)}]{Brinkmann00}
{Brinkmann}, W. \& {Kawai}, N. 2000, \aap, 363, 640

\bibitem[{{Brinkmann} {et~al.}(2005){Brinkmann}, {Kotani}, \&
  {Kawai}}]{Brinkmann05}
{Brinkmann}, W., {Kotani}, T., \& {Kawai}, N. 2005, \aap, 431, 575

\bibitem[{{Brown} {et~al.}(1991){Brown}, {Cassinelli}, \& {Collins}}]{Brown91}
{Brown}, J.~C., {Cassinelli}, J.~P., \& {Collins}, II, G.~W. 1991, \apj, 378,
  307

\bibitem[{{Cardelli} {et~al.}(1989){Cardelli}, {Clayton}, \&
  {Mathis}}]{Cardelli89}
{Cardelli}, J.~A., {Clayton}, G.~C., \& {Mathis}, J.~S. 1989, \apj, 345, 245

\bibitem[{{Clark} \& {Murdin}(1978)}]{Clark78}
{Clark}, D.~H. \& {Murdin}, P. 1978, \nat, 276, 44

\bibitem[{{Davidson} \& {McCray}(1980)}]{Davidson80}
{Davidson}, K. \& {McCray}, R. 1980, \apj, 241, 1082

\bibitem[{{Dolan} {et~al.}(1997){Dolan}, {Boyd}, {Fabrika}, {Tapia}, {Bychkov},
  {Panferov}, {Nelson}, {Percival}, {van Citters}, {Taylor}, \&
  {Taylor}}]{Dolan97}
{Dolan}, J.~F., {Boyd}, P.~T., {Fabrika}, S., {et~al.} 1997, \aap, 327, 648

\bibitem[{{Eikenberry} {et~al.}(2001){Eikenberry}, {Cameron}, {Fierce}, {Kull},
  {Dror}, {Houck}, \& {Margon}}]{Eikenberry01}
{Eikenberry}, S.~S., {Cameron}, P.~B., {Fierce}, B.~W., {et~al.} 2001, \apj,
  561, 1027

\bibitem[{{Fabian} \& {Rees}(1979)}]{Fabian79}
{Fabian}, A.~C. \& {Rees}, M.~J. 1979, \mnras, 187, 13P

\bibitem[{{Fabrika}(2004)}]{Fabrika04}
{Fabrika}, S. 2004, Astrophysics and Space Physics Reviews, 12, 1

\bibitem[{{Fabrika} {et~al.}(2015){Fabrika}, {Ueda}, {Vinokurov}, {Sholukhova},
  \& {Shidatsu}}]{Fabrika15}
{Fabrika}, S., {Ueda}, Y., {Vinokurov}, A., {Sholukhova}, O., \& {Shidatsu}, M.
  2015, Nature Physics, 11, 551

\bibitem[{{Fabrika} \& {Borisov}(1987)}]{Fabrika87}
{Fabrika}, S.~N. \& {Borisov}, N.~V. 1987, Soviet Astronomy Letters, 13, 279

\bibitem[{{Fabrika} \& {Sholukhova}(2008)}]{Fabrika08}
{Fabrika}, S.~N. \& {Sholukhova}, O. 2008, in Microquasars and Beyond, 52

\bibitem[{{Ferland} {et~al.}(2017){Ferland}, {Chatzikos}, {Guzm{\'a}n},
  {Lykins}, {van Hoof}, {Williams}, {Abel}, {Badnell}, {Keenan}, {Porter}, \&
  {Stancil}}]{Ferland17}
{Ferland}, G.~J., {Chatzikos}, M., {Guzm{\'a}n}, F., {et~al.} 2017, \rmxaa, 53,
  385

\bibitem[{{Ferland} {et~al.}(2013){Ferland}, {Porter}, {van Hoof}, {Williams},
  {Abel}, {Lykins}, {Shaw}, {Henney}, \& {Stancil}}]{Ferland13}
{Ferland}, G.~J., {Porter}, R.~L., {van Hoof}, P.~A.~M., {et~al.} 2013, \rmxaa,
  49, 137

\bibitem[{{Friedman} {et~al.}(2011){Friedman}, {York}, {McCall}, {Dahlstrom},
  {Sonnentrucker}, {Welty}, {Drosback}, {Hobbs}, {Rachford}, \&
  {Snow}}]{Friedman11}
{Friedman}, S.~D., {York}, D.~G., {McCall}, B.~J., {et~al.} 2011, \apj, 727, 33

\bibitem[{{Fuchs} {et~al.}(2006){Fuchs}, {Koch Miramond}, \&
  {{\'A}brah{\'a}m}}]{Fuchs06}
{Fuchs}, Y., {Koch Miramond}, L., \& {{\'A}brah{\'a}m}, P. 2006, \aap, 445,
  1041

\bibitem[{{Goranskii} {et~al.}(1998){Goranskii}, {Esipov}, \&
  {Cherepashchuk}}]{Goranskii98}
{Goranskii}, V.~P., {Esipov}, V.~F., \& {Cherepashchuk}, A.~M. 1998, Astronomy
  Reports, 42, 209

\bibitem[{{Gravity Collaboration} {et~al.}(2017{\natexlab{a}}){Gravity
  Collaboration}, {Abuter}, {Accardo}, {Amorim}, {Anugu}, {{\'A}vila},
  {Azouaoui}, {Benisty}, {Berger}, {Blind}, {Bonnet}, {Bourget}, {Brandner},
  {Brast}, {Buron}, {Burtscher}, {Cassaing}, {Chapron}, {Choquet},
  {Cl{\'e}net}, {Collin}, {Coud{\'e} Du Foresto}, {de Wit}, {de Zeeuw}, {Deen},
  {Delplancke-Str{\"o}bele}, {Dembet}, {Derie}, {Dexter}, {Duvert}, {Ebert},
  {Eckart}, {Eisenhauer}, {Esselborn}, {F{\'e}dou}, {Finger}, {Garcia}, {Garcia
  Dabo}, {Garcia Lopez}, {Gendron}, {Genzel}, {Gillessen}, {Gonte}, {Gordo},
  {Grould}, {Gr{\"o}zinger}, {Guieu}, {Haguenauer}, {Hans}, {Haubois}, {Haug},
  {Haussmann}, {Henning}, {Hippler}, {Horrobin}, {Huber}, {Hubert}, {Hubin},
  {Hummel}, {Jakob}, {Janssen}, {Jochum}, {Jocou}, {Kaufer}, {Kellner},
  {Kendrew}, {Kern}, {Kervella}, {Kiekebusch}, {Klein}, {Kok}, {Kolb}, {Kulas},
  {Lacour}, {Lapeyr{\`e}re}, {Lazareff}, {Le Bouquin}, {L{\`e}na}, {Lenzen},
  {L{\'e}v{\^e}que}, {Lippa}, {Magnard}, {Mehrgan}, {Mellein}, {M{\'e}rand},
  {Moreno-Ventas}, {Moulin}, {M{\"u}ller}, {M{\"u}ller}, {Neumann}, {Oberti},
  {Ott}, {Pallanca}, {Panduro}, {Pasquini}, {Paumard}, {Percheron}, {Perraut},
  {Perrin}, {Pfl{\"u}ger}, {Pfuhl}, {Phan Duc}, {Plewa}, {Popovic}, {Rabien},
  {Ram{\'{\i}}rez}, {Ramos}, {Rau}, {Riquelme}, {Rohloff}, {Rousset},
  {Sanchez-Bermudez}, {Scheithauer}, {Sch{\"o}ller}, {Schuhler}, {Spyromilio},
  {Straubmeier}, {Sturm}, {Suarez}, {Tristram}, {Ventura}, {Vincent},
  {Waisberg}, {Wank}, {Weber}, {Wieprecht}, {Wiest}, {Wiezorrek}, {Wittkowski},
  {Woillez}, {Wolff}, {Yazici}, {Ziegler}, \& {Zins}}]{GRAVITY17}
{Gravity Collaboration}, {Abuter}, R., {Accardo}, M., {et~al.}
  2017{\natexlab{a}}, \aap, 602, A94

\bibitem[{{Gravity Collaboration} {et~al.}(2017{\natexlab{b}}){Gravity
  Collaboration}, {Petrucci}, {Waisberg}, {Le Bouquin}, {Dexter}, {Dubus},
  {Perraut}, {Kervella}, {Abuter}, {Amorim}, {Anugu}, {Berger}, {Blind},
  {Bonnet}, {Brandner}, {Buron}, {Choquet}, {Cl{\'e}net}, {de Wit}, {Deen},
  {Eckart}, {Eisenhauer}, {Finger}, {Garcia}, {Garcia Lopez}, {Gendron},
  {Genzel}, {Gillessen}, {Gonte}, {Haubois}, {Haug}, {Haussmann}, {Henning},
  {Hippler}, {Horrobin}, {Hubert}, {Jochum}, {Jocou}, {Kok}, {Kolb}, {Kulas},
  {Lacour}, {Lazareff}, {L{\`e}na}, {Lippa}, {M{\'e}rand}, {M{\"u}ller}, {Ott},
  {Panduro}, {Paumard}, {Perrin}, {Pfuhl}, {Ramos}, {Rau}, {Rohloff},
  {Rousset}, {Sanchez-Bermudez}, {Scheithauer}, {Sch{\"o}ller}, {Straubmeier},
  {Sturm}, {Vincent}, {Wank}, {Wieprecht}, {Wiest}, {Wiezorrek}, {Wittkowski},
  {Woillez}, {Yazici}, \& {Zins}}]{GRAVITYSS43317}
{Gravity Collaboration}, {Petrucci}, P.-O., {Waisberg}, I., {et~al.}
  2017{\natexlab{b}}, \aap, 602, L11

\bibitem[{{Green} {et~al.}(2018){Green}, {Schlafly}, {Finkbeiner}, {Rix},
  {Martin}, {Burgett}, {Draper}, {Flewelling}, {Hodapp}, {Kaiser}, {Kudritzki},
  {Magnier}, {Metcalfe}, {Tonry}, {Wainscoat}, \& {Waters}}]{Green18}
{Green}, G.~M., {Schlafly}, E.~F., {Finkbeiner}, D., {et~al.} 2018, \mnras,
  478, 651

\bibitem[{{Herbig}(1975)}]{Herbig75}
{Herbig}, G.~H. 1975, \apj, 196, 129

\bibitem[{{Hjellming} \& {Johnston}(1981)}]{Hjellming81}
{Hjellming}, R.~M. \& {Johnston}, K.~J. 1981, \apjl, 246, L141

\bibitem[{Hunter(2007)}]{Hunter2007}
Hunter, J.~D. 2007, Computing In Science \& Engineering, 9, 90

\bibitem[{{Kaaret} {et~al.}(2017){Kaaret}, {Feng}, \& {Roberts}}]{Kaaret17}
{Kaaret}, P., {Feng}, H., \& {Roberts}, T.~P. 2017, \araa, 55, 303

\bibitem[{{Kausch} {et~al.}(2015){Kausch}, {Noll}, {Smette}, {Kimeswenger},
  {Barden}, {Szyszka}, {Jones}, {Sana}, {Horst}, \& {Kerber}}]{Kausch15}
{Kausch}, W., {Noll}, S., {Smette}, A., {et~al.} 2015, \aap, 576, A78

\bibitem[{{Kos} \& {Zwitter}(2013)}]{Kos13}
{Kos}, J. \& {Zwitter}, T. 2013, \apj, 774, 72

\bibitem[{{Kotani} {et~al.}(1996){Kotani}, {Kawai}, {Matsuoka}, \&
  {Brinkmann}}]{Kotani96}
{Kotani}, T., {Kawai}, N., {Matsuoka}, M., \& {Brinkmann}, W. 1996, \pasj, 48,
  619

\bibitem[{{Lampton} {et~al.}(1976){Lampton}, {Margon}, \& {Bowyer}}]{Lampton76}
{Lampton}, M., {Margon}, B., \& {Bowyer}, S. 1976, \apj, 208, 177

\bibitem[{{Liu} {et~al.}(2015){Liu}, {Bai}, {Wang}, {Justham}, {Lu}, {Gu},
  {Liu}, {di Stefano}, {Guo}, {Cabrera-Lavers}, {{\'A}lvarez}, {Cao}, \&
  {Kulkarni}}]{Liu15}
{Liu}, J.-F., {Bai}, Y., {Wang}, S., {et~al.} 2015, \nat, 528, 108

\bibitem[{{Lockman} {et~al.}(2007){Lockman}, {Blundell}, \& {Goss}}]{Lockman07}
{Lockman}, F.~J., {Blundell}, K.~M., \& {Goss}, W.~M. 2007, \mnras, 381, 881

\bibitem[{{Lopez} {et~al.}(2014){Lopez}, {Lagarde}, {Jaffe}, {Petrov},
  {Sch{\"o}ller}, {Antonelli}, {Beckmann}, {Berio}, {Bettonvil}, {Glindemann},
  {Gonzalez}, {Graser}, {Hofmann}, {Millour}, {Robbe-Dubois}, {Venema}, {Wolf},
  {Henning}, {Lanz}, {Weigelt}, {Agocs}, {Bailet}, {Bresson}, {Bristow},
  {Dugu{\'e}}, {Heininger}, {Kroes}, {Laun}, {Lehmitz}, {Neumann}, {Augereau},
  {Avila}, {Behrend}, {van Belle}, {Berger}, {van Boekel}, {Bonhomme},
  {Bourget}, {Brast}, {Clausse}, {Connot}, {Conzelmann}, {Cruzal{\`e}bes},
  {Csepany}, {Danchi}, {Delbo}, {Delplancke}, {Dominik}, {van Duin}, {Elswijk},
  {Fantei}, {Finger}, {Gabasch}, {Gay}, {Girard}, {Girault}, {Gitton},
  {Glazenborg}, {Gont{\'e}}, {Guitton}, {Guniat}, {De Haan}, {Haguenauer},
  {Hanenburg}, {Hogerheijde}, {ter Horst}, {Hron}, {Hugues}, {Hummel},
  {Idserda}, {Ives}, {Jakob}, {Jasko}, {Jolley}, {Kiraly}, {K{\"o}hler},
  {Kragt}, {Kroener}, {Kuindersma}, {Labadie}, {Leinert}, {Le Poole}, {Lizon},
  {Lucuix}, {Marcotto}, {Martinache}, {Martinot-Lagarde}, {Mathar}, {Matter},
  {Mauclert}, {Mehrgan}, {Meilland}, {Meisenheimer}, {Meisner}, {Mellein},
  {Menardi}, {Menut}, {Merand}, {Morel}, {Mosoni}, {Navarro}, {Nussbaum},
  {Ottogalli}, {Palsa}, {Panduro}, {Pantin}, {Parra}, {Percheron}, {Duc},
  {Pott}, {Pozna}, {Przygodda}, {Rabbia}, {Richichi}, {Rigal}, {Roelfsema},
  {Rupprecht}, {Schertl}, {Schmidt}, {Schuhler}, {Schuil}, {Spang},
  {Stegmeier}, {Thiam}, {Tromp}, {Vakili}, {Vannier}, {Wagner}, \&
  {Woillez}}]{Lopez14}
{Lopez}, B., {Lagarde}, S., {Jaffe}, W., {et~al.} 2014, The Messenger, 157, 5

\bibitem[{{Luri} {et~al.}(2018){Luri}, {Brown}, {Sarro}, {Arenou},
  {Bailer-Jones}, {Castro-Ginard}, {de Bruijne}, {Prusti}, {Babusiaux}, \&
  {Delgado}}]{Luri18}
{Luri}, X., {Brown}, A.~G.~A., {Sarro}, L.~M., {et~al.} 2018, \aap, 616, A9

\bibitem[{{Margon}(1984)}]{Margon84}
{Margon}, B. 1984, \araa, 22, 507

\bibitem[{{Margon} {et~al.}(1979){Margon}, {Ford}, {Grandi}, \&
  {Stone}}]{Margon79}
{Margon}, B., {Ford}, H.~C., {Grandi}, S.~A., \& {Stone}, R.~P.~S. 1979, \apjl,
  233, L63

\bibitem[{{Marshall} {et~al.}(2013){Marshall}, {Canizares}, {Hillwig},
  {Mioduszewski}, {Rupen}, {Schulz}, {Nowak}, \& {Heinz}}]{Marshall13}
{Marshall}, H.~L., {Canizares}, C.~R., {Hillwig}, T., {et~al.} 2013, \apj, 775,
  75

\bibitem[{{Marshall} {et~al.}(2002){Marshall}, {Canizares}, \&
  {Schulz}}]{Marshall02}
{Marshall}, H.~L., {Canizares}, C.~R., \& {Schulz}, N.~S. 2002, \apj, 564, 941

\bibitem[{{Medvedev} \& {Fabrika}(2010)}]{Medvedev10}
{Medvedev}, A. \& {Fabrika}, S. 2010, \mnras, 402, 479

\bibitem[{{Middleton} {et~al.}(2018){Middleton}, {Walton}, {Alston}, {Dauser},
  {Eikenberry}, {Jiang}, {Fabian}, {Fuerst}, {Brightman}, {Marshall}, {Parker},
  {Pinto}, {Harrison}, {Bachetti}, {Altamirano}, {Bird}, {Perez},
  {Miller-Jones}, {Charles}, {Boggs}, {Christensen}, {Craig}, {Forster},
  {Grefenstette}, {Hailey}, {Madsen}, {Stern}, \& {Zhang}}]{Middleton18}
{Middleton}, M.~J., {Walton}, D.~J., {Alston}, W., {et~al.} 2018, ArXiv
  e-prints [\eprint[arXiv]{1810.10518}]

\bibitem[{{Murdin} {et~al.}(1980){Murdin}, {Clark}, \& {Martin}}]{Murdin80}
{Murdin}, P., {Clark}, D.~H., \& {Martin}, P.~G. 1980, \mnras, 193, 135

\bibitem[{{Osterbrock} \& {Ferland}(2006)}]{AGN3}
{Osterbrock}, D.~E. \& {Ferland}, G.~J. 2006, {Astrophysics of gaseous nebulae
  and active galactic nuclei}

\bibitem[{{Panferov} \& {Fabrika}(1993)}]{Panferov93}
{Panferov}, A.~A. \& {Fabrika}, S.~N. 1993, Astronomy Letters, 19, 41

\bibitem[{{Panferov} \& {Fabrika}(1997)}]{Panferov97a}
{Panferov}, A.~A. \& {Fabrika}, S.~N. 1997, Astronomy Reports, 41, 506

\bibitem[{{Panferov} {et~al.}(1997){Panferov}, {Fabrika}, \&
  {Rakhimov}}]{Panferov97b}
{Panferov}, A.~A., {Fabrika}, S.~N., \& {Rakhimov}, V.~Y. 1997, \azh, 74, 392

\bibitem[{{Paragi} {et~al.}(1999){Paragi}, {Vermeulen}, {Fejes}, {Schilizzi},
  {Spencer}, \& {Stirling}}]{Paragi99}
{Paragi}, Z., {Vermeulen}, R.~C., {Fejes}, I., {et~al.} 1999, \aap, 348, 910

\bibitem[{{Perez M.} \& {Blundell}(2010)}]{Perez10}
{Perez M.}, S. \& {Blundell}, K.~M. 2010, \mnras, 408, 2

\bibitem[{{Roberts}(1974)}]{Roberts74}
{Roberts}, W.~J. 1974, \apj, 187, 575

\bibitem[{{Schlegel} {et~al.}(1998){Schlegel}, {Finkbeiner}, \&
  {Davis}}]{Schlegel98}
{Schlegel}, D.~J., {Finkbeiner}, D.~P., \& {Davis}, M. 1998, \apj, 500, 525

\bibitem[{{Shakura} \& {Sunyaev}(1973)}]{Shakura73}
{Shakura}, N.~I. \& {Sunyaev}, R.~A. 1973, \aap, 24, 337

\bibitem[{{Smette} {et~al.}(2015){Smette}, {Sana}, {Noll}, {Horst}, {Kausch},
  {Kimeswenger}, {Barden}, {Szyszka}, {Jones}, {Gallenne}, {Vinther},
  {Ballester}, \& {Taylor}}]{Smette15}
{Smette}, A., {Sana}, H., {Noll}, S., {et~al.} 2015, \aap, 576, A77

\bibitem[{{Stephenson} \& {Sanduleak}(1977)}]{Stephenson77}
{Stephenson}, C.~B. \& {Sanduleak}, N. 1977, \apjs, 33, 459

\bibitem[{{Stirling} {et~al.}(2002){Stirling}, {Jowett}, {Spencer}, {Paragi},
  {Ogley}, \& {Cawthorne}}]{Stirling02}
{Stirling}, A.~M., {Jowett}, F.~H., {Spencer}, R.~E., {et~al.} 2002, \mnras,
  337, 657

\bibitem[{{Tody}(1986)}]{Tody86}
{Tody}, D. 1986, in \procspie, Vol. 627, Instrumentation in astronomy VI, ed.
  D.~L. {Crawford}, 733

\bibitem[{{Tody}(1993)}]{Tody93}
{Tody}, D. 1993, in Astronomical Society of the Pacific Conference Series,
  Vol.~52, Astronomical Data Analysis Software and Systems II, ed. R.~J.
  {Hanisch}, R.~J.~V. {Brissenden}, \& J.~{Barnes}, 173

\bibitem[{{van den Heuvel} {et~al.}(1980){van den Heuvel}, {Ostriker}, \&
  {Petterson}}]{vandenHeuvel80}
{van den Heuvel}, E.~P.~J., {Ostriker}, J.~P., \& {Petterson}, J.~A. 1980,
  \aap, 81, L7

\bibitem[{Van Der~Walt {et~al.}(2011)Van Der~Walt, Colbert, \&
  Varoquaux}]{van2011numpy}
Van Der~Walt, S., Colbert, S.~C., \& Varoquaux, G. 2011, Computing in Science
  \& Engineering, 13, 22

\bibitem[{{Vermeulen} {et~al.}(1993){Vermeulen}, {Murdin}, {van den Heuvel},
  {Fabrika}, {Wagner}, {Margon}, {Hutchings}, {Schilizzi}, {van Kerkwijk}, {van
  den Hoek}, {Ott}, {Angebault}, {Miley}, {D'Odorico}, \&
  {Borisov}}]{Vermeulen93}
{Vermeulen}, R.~C., {Murdin}, P.~G., {van den Heuvel}, E.~P.~J., {et~al.} 1993,
  \aap, 270, 204

\bibitem[{{Vernet} {et~al.}(2011){Vernet}, {Dekker}, {D'Odorico}, {Kaper},
  {Kjaergaard}, {Hammer}, {Randich}, {Zerbi}, {Groot}, {Hjorth}, {Guinouard},
  {Navarro}, {Adolfse}, {Albers}, {Amans}, {Andersen}, {Andersen}, {Binetruy},
  {Bristow}, {Castillo}, {Chemla}, {Christensen}, {Conconi}, {Conzelmann},
  {Dam}, {de Caprio}, {de Ugarte Postigo}, {Delabre}, {di Marcantonio},
  {Downing}, {Elswijk}, {Finger}, {Fischer}, {Flores}, {Fran{\c c}ois},
  {Goldoni}, {Guglielmi}, {Haigron}, {Hanenburg}, {Hendriks}, {Horrobin},
  {Horville}, {Jessen}, {Kerber}, {Kern}, {Kiekebusch}, {Kleszcz}, {Klougart},
  {Kragt}, {Larsen}, {Lizon}, {Lucuix}, {Mainieri}, {Manuputy}, {Martayan},
  {Mason}, {Mazzoleni}, {Michaelsen}, {Modigliani}, {Moehler}, {M{\o}ller},
  {Norup S{\o}rensen}, {N{\o}rregaard}, {P{\'e}roux}, {Patat}, {Pena}, {Pragt},
  {Reinero}, {Rigal}, {Riva}, {Roelfsema}, {Royer}, {Sacco}, {Santin},
  {Schoenmaker}, {Spano}, {Sweers}, {Ter Horst}, {Tintori}, {Tromp}, {van
  Dael}, {van der Vliet}, {Venema}, {Vidali}, {Vinther}, {Vola}, {Winters},
  {Wistisen}, {Wulterkens}, \& {Zacchei}}]{Vernet11}
{Vernet}, J., {Dekker}, H., {D'Odorico}, S., {et~al.} 2011, \aap, 536, A105

\bibitem[{{Wagner}(1986)}]{Wagner86}
{Wagner}, R.~M. 1986, \apj, 308, 152

\bibitem[{{Watson} {et~al.}(1986){Watson}, {Stewart}, {Brinkmann}, \&
  {King}}]{Watson86}
{Watson}, M.~G., {Stewart}, G.~C., {Brinkmann}, W., \& {King}, A.~R. 1986,
  \mnras, 222, 261

\bibitem[{{Whitmire} \& {Matese}(1980)}]{Whitmire80}
{Whitmire}, D.~P. \& {Matese}, J.~J. 1980, \mnras, 193, 707

\bibitem[{{Zealey} {et~al.}(1980){Zealey}, {Dopita}, \& {Malin}}]{Zealy80}
{Zealey}, W.~J., {Dopita}, M.~A., \& {Malin}, D.~F. 1980, \mnras, 192, 731

\end{thebibliography}

\begin{appendix}
\onecolumn 

\section{GRAVITY: Full Data and Model Fits} 

Here we show as in Figure \ref{fig:model_plot_P99_3} the full K band spectrum and interferometric data, as well as the best fit jet models, for the remaining GRAVITY observations. Compact and extended jet knots are shown in colored full and dashed lines, respectively. Stationary emission lines are marked in green. In the 2017 observations, in which there is strong blending between the jets and the Br$\gamma$ stationary line, a combined fit is done, but we show only the jet model for clarity. For the Br$\gamma$ stationary line model and results, we refer to the companion paper on the equatorial outflows (Waisberg et al., sub.). 

\begin{figure*}[tb]
\label{fig:model_plot_COM} 
\centering
\includegraphics[width=1.0\columnwidth]{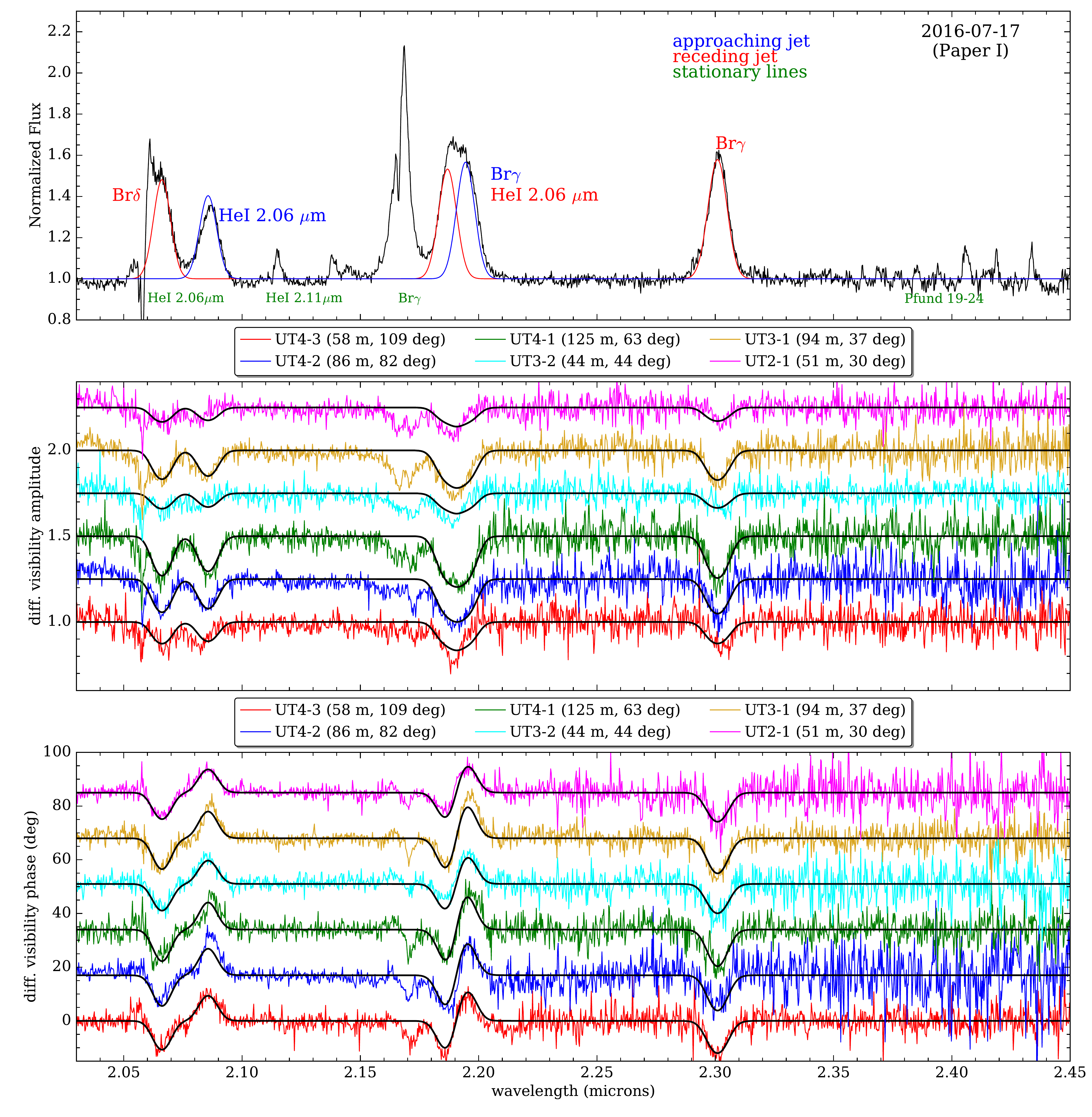}
\caption{Data and best fit jet model for the 2016 GRAVITY observation.}
\end{figure*}

\begin{figure*}[tb]
\label{fig:model_plot_P99_1} 
\centering
\includegraphics[width=1.0\columnwidth]{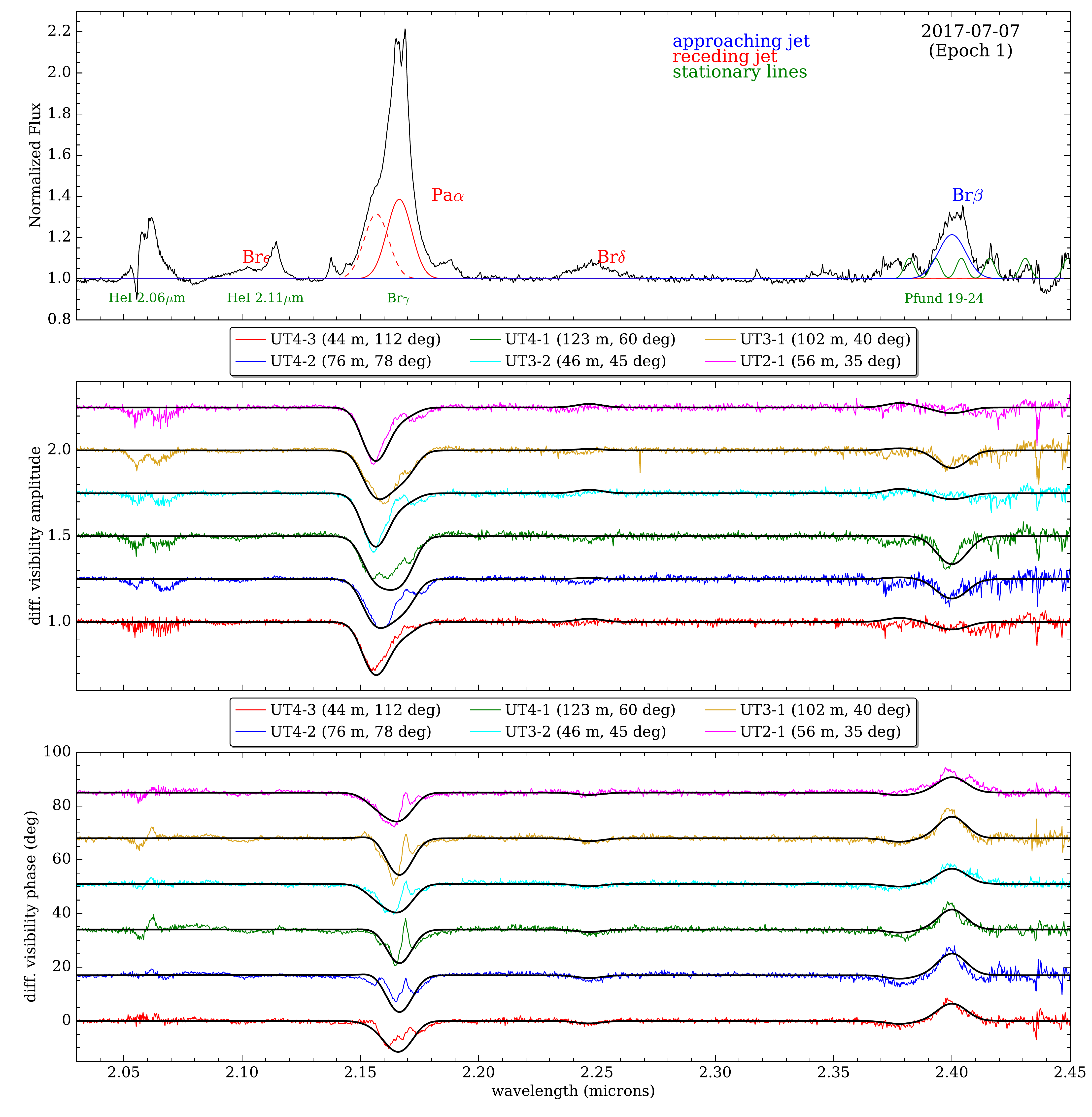}
\caption{Data and best fit jet model for the 2017 Epoch 1 GRAVITY observation.}
\end{figure*}

\begin{figure*}[tb]
\label{fig:model_plot_P99_2} 
\centering
\includegraphics[width=1.0\columnwidth]{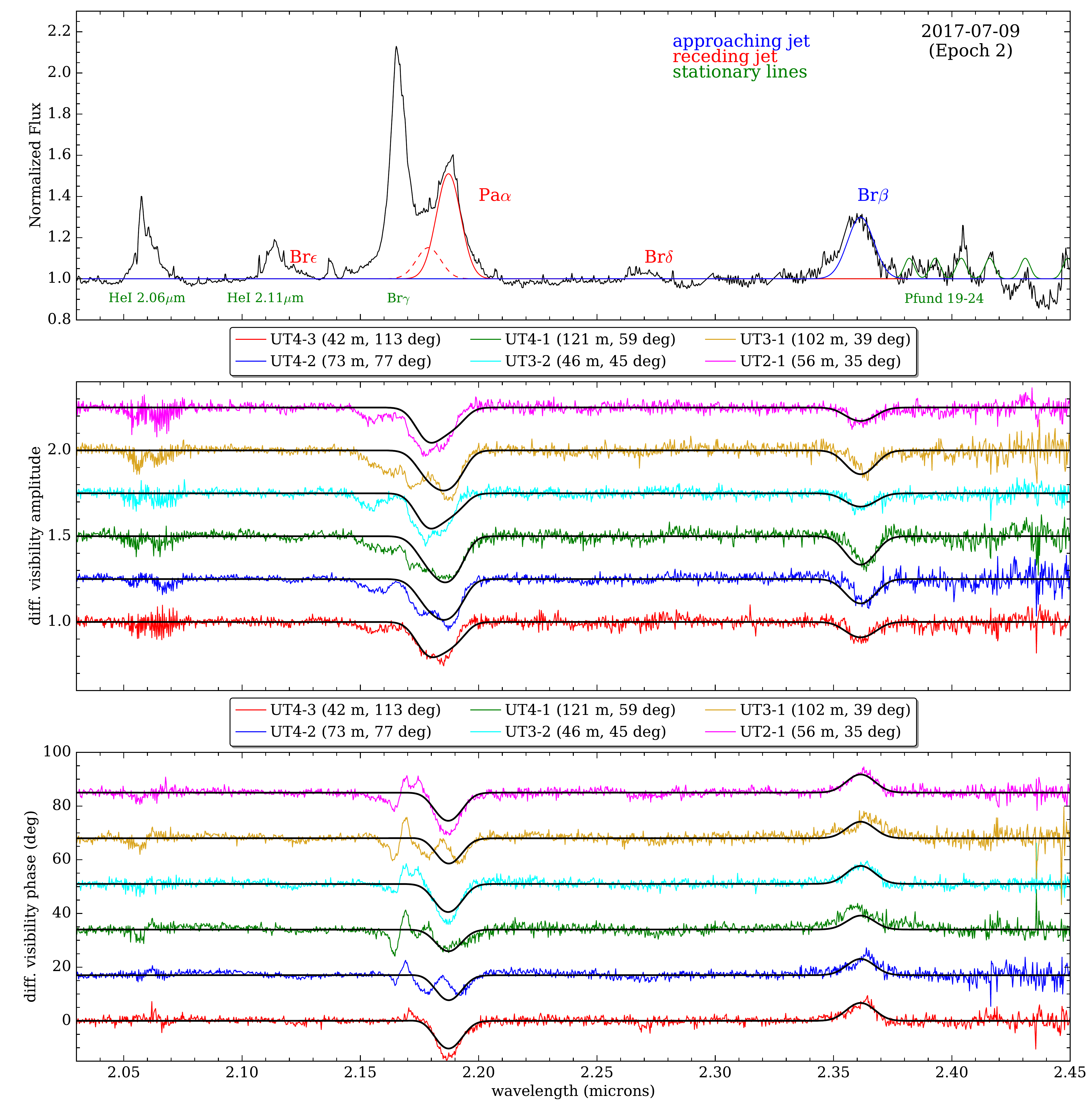}
\caption{Data and best fit jet model for the 2017 Epoch 2 GRAVITY observation.}
\end{figure*}

\section{XSHOOTER: Full Data and Model Fits} 

Here we show the data and best fits for the remaining XSHOOTER epochs as in Figure \ref{fig:model_XSHOOTER_1}. 

\begin{figure*}[tb]
\centering
\includegraphics[width=1.0\columnwidth]{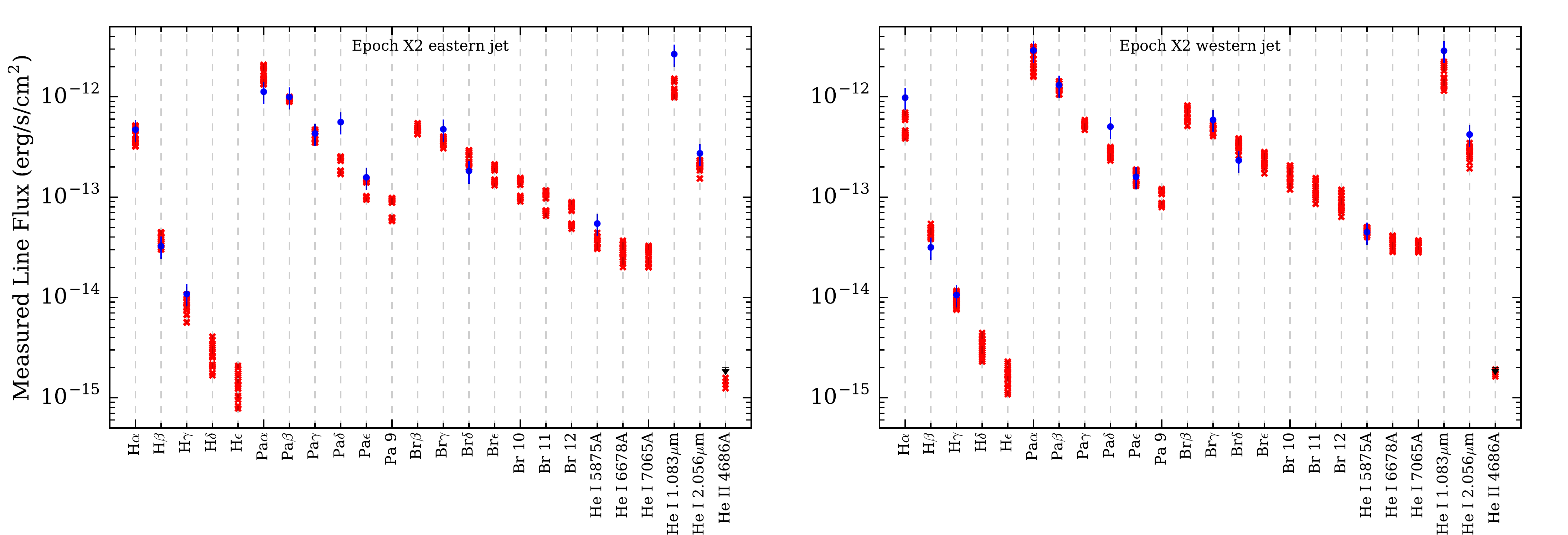} 
\caption{Data and best fit models for Epoch 2 of the XSHOOTER observations.}
\label{fig:model_XSHOOTER_2}
\end{figure*}

\begin{figure*}[tb]
\centering
\includegraphics[width=1.0\columnwidth]{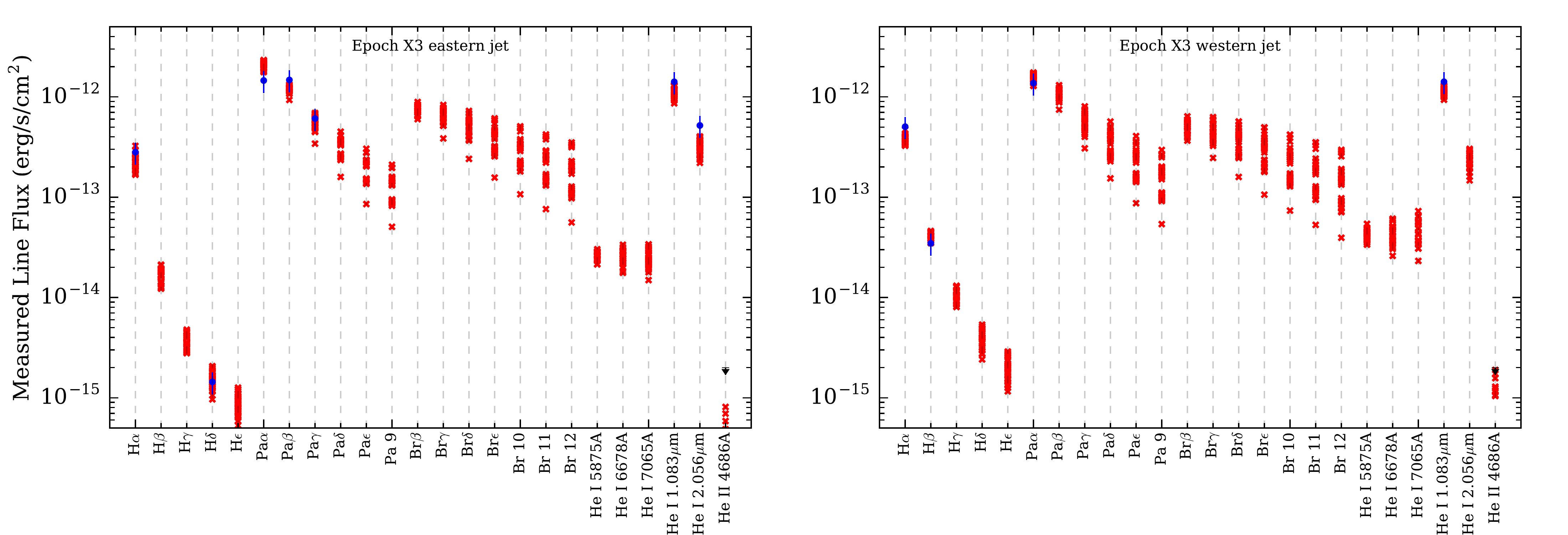} 
\caption{Data and best fit models for Epoch 3 of the XSHOOTER observations.}
\label{fig:model_XSHOOTER_3}
\end{figure*}

\begin{figure*}[tb]
\centering
\includegraphics[width=1.0\columnwidth]{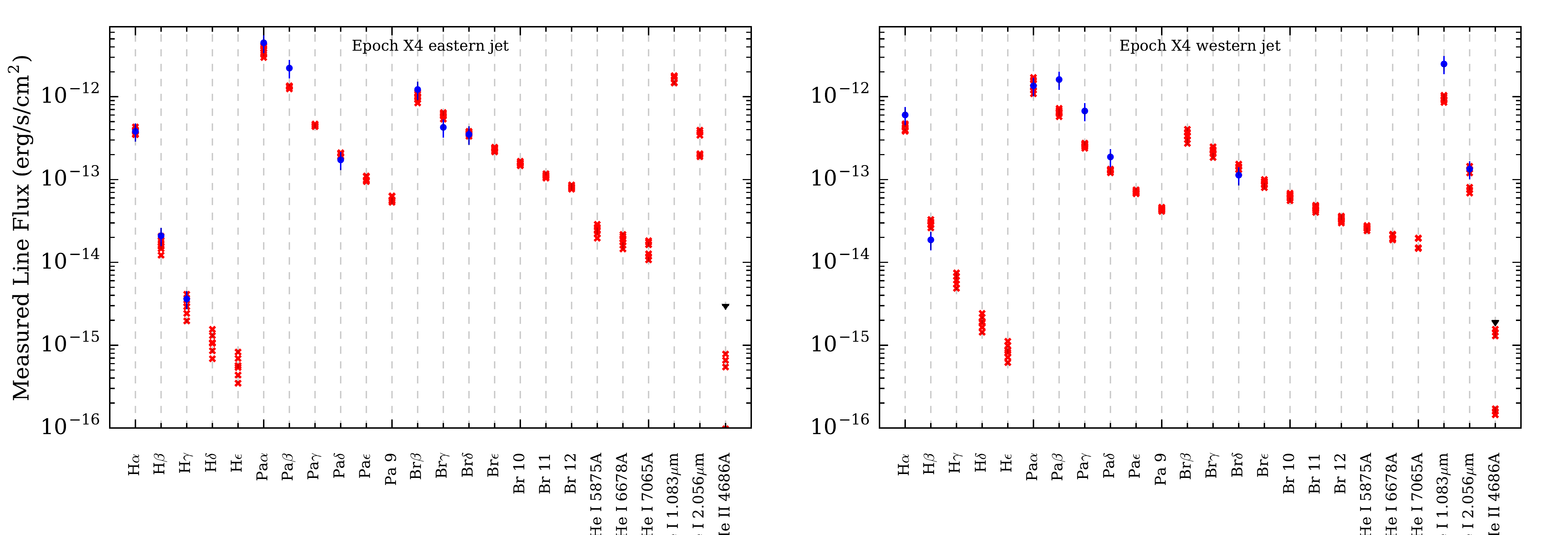} 
\caption{Data and best fit models for Epoch 4 of the XSHOOTER observations.}
\label{fig:model_XSHOOTER_4}
\end{figure*}

\begin{figure*}[tb]
\centering
\includegraphics[width=1.0\columnwidth]{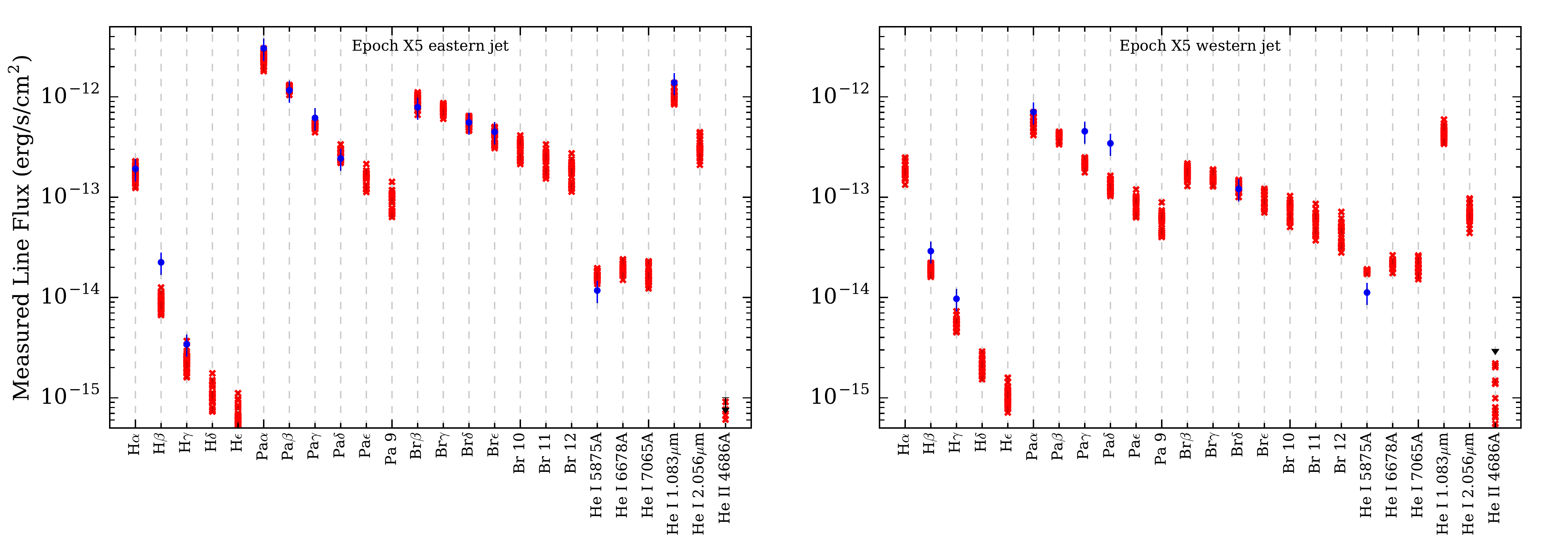} 
\caption{Data and best fit models for Epoch 5 of the XSHOOTER observations.}
\label{fig:model_XSHOOTER_5}
\end{figure*}

\end{appendix} 

\end{document}